\pdfoutput=1
\documentclass[conference]{IEEEtran}
\IEEEoverridecommandlockouts

\usepackage{cite}
\usepackage{amsmath,amssymb,amsfonts}
\usepackage{algorithmic}
\usepackage{graphicx}
\usepackage{caption}
\usepackage{subcaption}
\usepackage{textcomp}
\usepackage{xcolor}
\def\BibTeX{{\rm B\kern-.05em{\sc i\kern-.025em b}\kern-.08em
    T\kern-.1667em\lower.7ex\hbox{E}\kern-.125emX}}
\begin{document}

\title{Systematic parameterization of heat-assisted magnetic recording switching probabilities and the consequences for the resulting SNR}

\author{\IEEEauthorblockN{Florian Slanovc}
\IEEEauthorblockA{\textit{Faculty of Physics} \\
\textit{University of Vienna}\\
Vienna, Austria \\
florian.slanovc@univie.ac.at}
\and
\IEEEauthorblockN{Christoph Vogler}
\IEEEauthorblockA{\textit{Faculty of Physics} \\
\textit{University of Vienna}\\
Vienna, Austria \\
christoph.vogler@univie.ac.at}
\and
\IEEEauthorblockN{Olivia Muthsam}
\IEEEauthorblockA{\textit{Faculty of Physics} \\
\textit{University of Vienna}\\
Vienna, Austria \\
olivia.muthsam@univie.ac.at}
\and
\IEEEauthorblockN{Dieter Suess}
\IEEEauthorblockA{\textit{Faculty of Physics} \\
\textit{University of Vienna}\\
Vienna, Austria \\
dieter.suess@univie.ac.at}
}

\maketitle

\begin{abstract}
The signal-to-noise ratio (SNR) of a bit series written with heat-assisted magnetic recording (HAMR)
on granular media depends on a large number of different parameters. The choice of material properties is essential for the
obtained switching probabilities of single grains and therefore for the written bits' quality in terms of SNR. Studies where the effects of different material compositions on transition jitter and the switching probability are evaluated were done, but it is not
obvious, how significant those improvements will finally change the received SNR. To investigate that influence, we
developed an analytical model of the switching probability phase diagram, which contains independent parameters for,
inter alia, transition width, switching probability and curvature. Different values lead to corresponding bit
patterns on granular media, where a reader model detects the resulting signal, which is finally converted to a parameter
dependent SNR value. For grain diameters between 4 and 8~nm, we show an increase of \texttildelow10~dB for bit lengths between 4 and 12~nm, an increase of \texttildelow9~dB for maximum switching probabilities between 0.64 and 1.00, a decrease of \texttildelow5~dB for down-track-jitter parameters between 0 and 4~nm and an increase of \texttildelow1~dB for reduced  bit curvature.  Those results are furthermore compared to the theoretical formulas for
the SNR. We obtain a good agreement, even though we show slight deviations.
\end{abstract}

\section{Introduction}
The quality of a written bit pattern on granular media is mostly determined by the signal-to-noise ratio (SNR). A high SNR means low noise and a sharp edge between neighboring bits. The SNR depends either on suitable magnetic material properties of the grains to provide a good switching probability in the writing process but also on the size and position distribution of the grains in the granular medium. There are different methods to calculate the switching probability of a grain model during heat-assisted magnetic recording (HAMR), which is subject to a heat pulse and an external magnetic field. One method is solving the stochastic Landau-Lifshitz-Gilbert (LLG) equation for each atom of the grain (Ref.~\cite{b11}), another is solving the stochastic coarse-grained Landau-Lifshiftz-Bloch (LLB) equation (Ref.~\cite{b1}). Depending on the down-track position $d$ and off-track position $y$, the repetition of switching trajectories for a given parameter set results in the approximate switching probability of a grain. Calculating the probability for various $d$ and $y$ yields a phase diagram of the writing process. Instead of the off-track direction $y$, the peak temperature $T_\text{peak}$ can be used, because then there is no need to specify the maximum temperature at the track center beforehand. For each peak temperature $T_{\text{peak}}$ the off-track direction $y$ can be easily determined under the assumption of a Gaussian heat pulse via the relation
\begin{align}\label{eq::peakofftrack}
 T_{\text{peak}}(y) = (T_{\max} - T_{\min} ) \cdot \exp{\left(-\frac{y^2}{2\sigma^2}\right)} + T_{\min},
\end{align}
where $T_{\min}$ and $T_{\max}$ are the overall minimum and maximum temperatures of the whole heat pulse and $\sigma=\text{FWHM} / \sqrt{8\ln{2}}$ its standard deviation.
The final phase diagrams for the switching probability $P(d,T_\text{peak})$ (as in Ref.~\cite[Fig.2]{b7}) allow to simulate writing processes of bit patterns on granular media as we will describe in Sec. \ref{sec::printing}. The advantage of this approach is that the phase diagram has to be created only once and the switching probability can afterwards be extracted for an arbitrary amount of magnetic grains with no further computational effort. The disadvantage is that every grain is regarded individually and therefore it is not obvious how to take stray-field interactions into account. This can be done by adjusting the intrinsic distribution of the Curie temperature as shown in Ref.~\cite{b12}. The phase plots contain much information about the size and characteristics of the magnetic grains as well as the parameters of the writing process (velocity, external applied field etc.). In Ref.~\cite{b2} for instance, the shape of the phase diagram depending on the composition of a bi-layer material is investigated. This information only refers to a single grain but a priori tells very little about the resulting SNR of the read-back signal of a bit series. Hence, there is need to systematically investigate the influence of changes of the phase diagram on the final SNR. In this work we will develop a mathematical model of a phase diagram, which allows to vary certain parameters and perform writing processes with the resulting diagrams. The read-back signal then gives some indication of the potential SNR-improvement. The mathematical formulation of the phase plot is presented in Sec. \ref{sec::model}. In Sec. \ref{sec::printing}, we will describe the simulation of the writing and read-back processes and Sec. \ref{sec::results} will summarize and discuss the results of the received SNR and compare those to theoretical formulas. 

\section{Mathematical model of a phase plot}\label{sec::model}
In heat-assisted magnetic recording (HAMR) the switching probability of a magnetic grain can be represented as a phase diagram. As in Ref.~\cite[Fig. 2]{b7} the area of the highest switching probability has a C-like shape in the $d-T_\text{peak}$-plane. In the following we define an analytical function $P(d,T)$, which allows to fit such a phase plot.

\subsection{Model parameters}\label{sec::modellparameters}
We use eight parameters that fully determine the shape of the phase plot: 
\begin{itemize}
\item down-track-jitter parameter: $\sigma_d$ [nm]
\item off-track-jitter parameter: $\sigma_o$ [K]
\item maximum switching probability: $P_\text{max}$
\item half maximum temperature: $F$ [K]
\item bit length: $b$ [nm]
\item curvature parameter: $p_1$ [nm/$\text{K}^2$]
\item position in $T_\text{peak}$-direction: $p_2$ [K]
\item position in $d$-direction: $p_3$ [nm]
\end{itemize}

\subsection{Mathematical model}
\begin{figure}[htbp]
 \begin{subfigure}{0.50\textwidth}
\centerline{\includegraphics[width=1.00\textwidth]{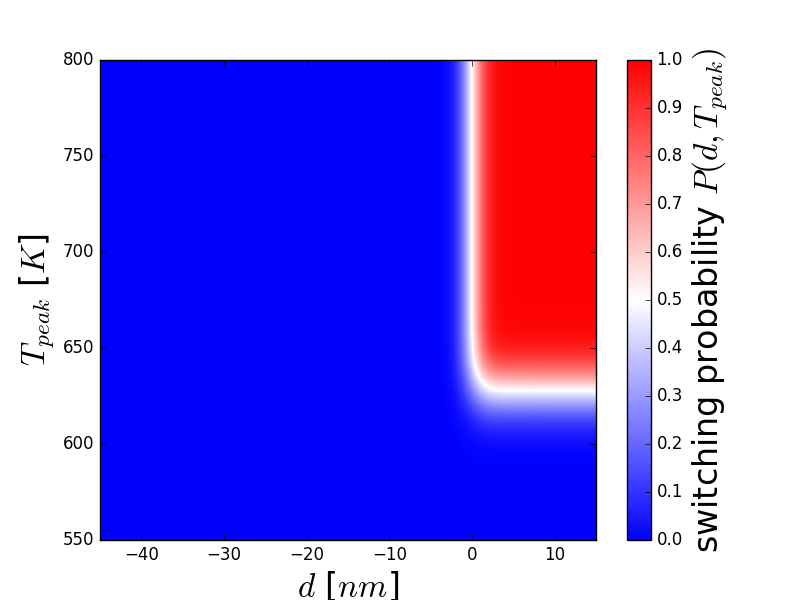}}
 \caption{}
 \end{subfigure}
  \begin{subfigure}{0.50\textwidth}
\centerline{\includegraphics[width=1.00\textwidth]{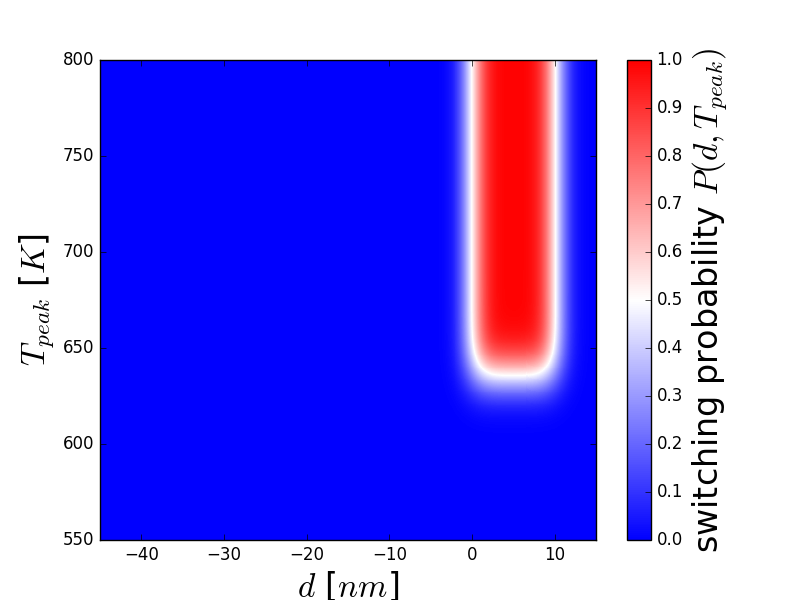}}
 \caption{}
 \end{subfigure}
   \begin{subfigure}{0.50\textwidth}
\centerline{\includegraphics[width=1.00\textwidth]{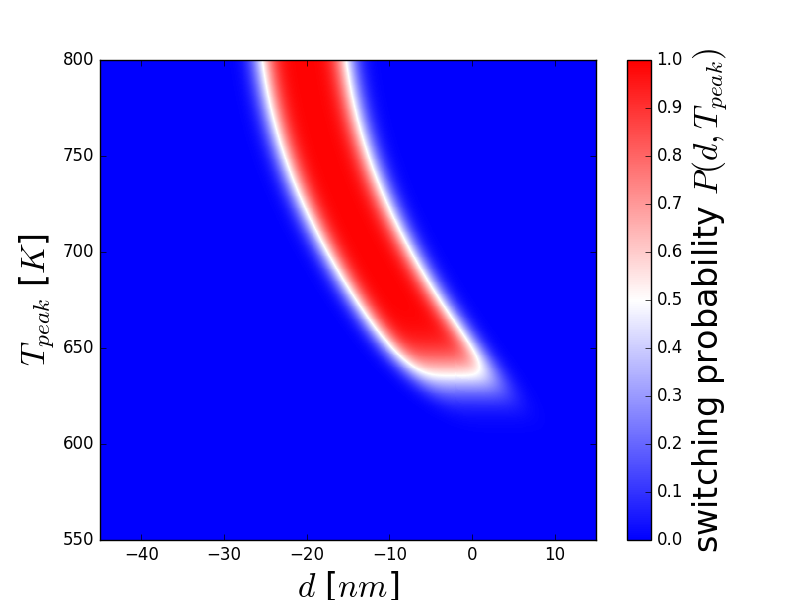}}
 \caption{}
 \end{subfigure}
\caption{(a) and (b) show the graphs of the help functions $h_1$ and $h_2$. In (c) the final model phase plot can be observed.}
\label{fig::progress}
\end{figure}

\begin{figure}[htbp]
\centerline{\includegraphics[width=0.50\textwidth]{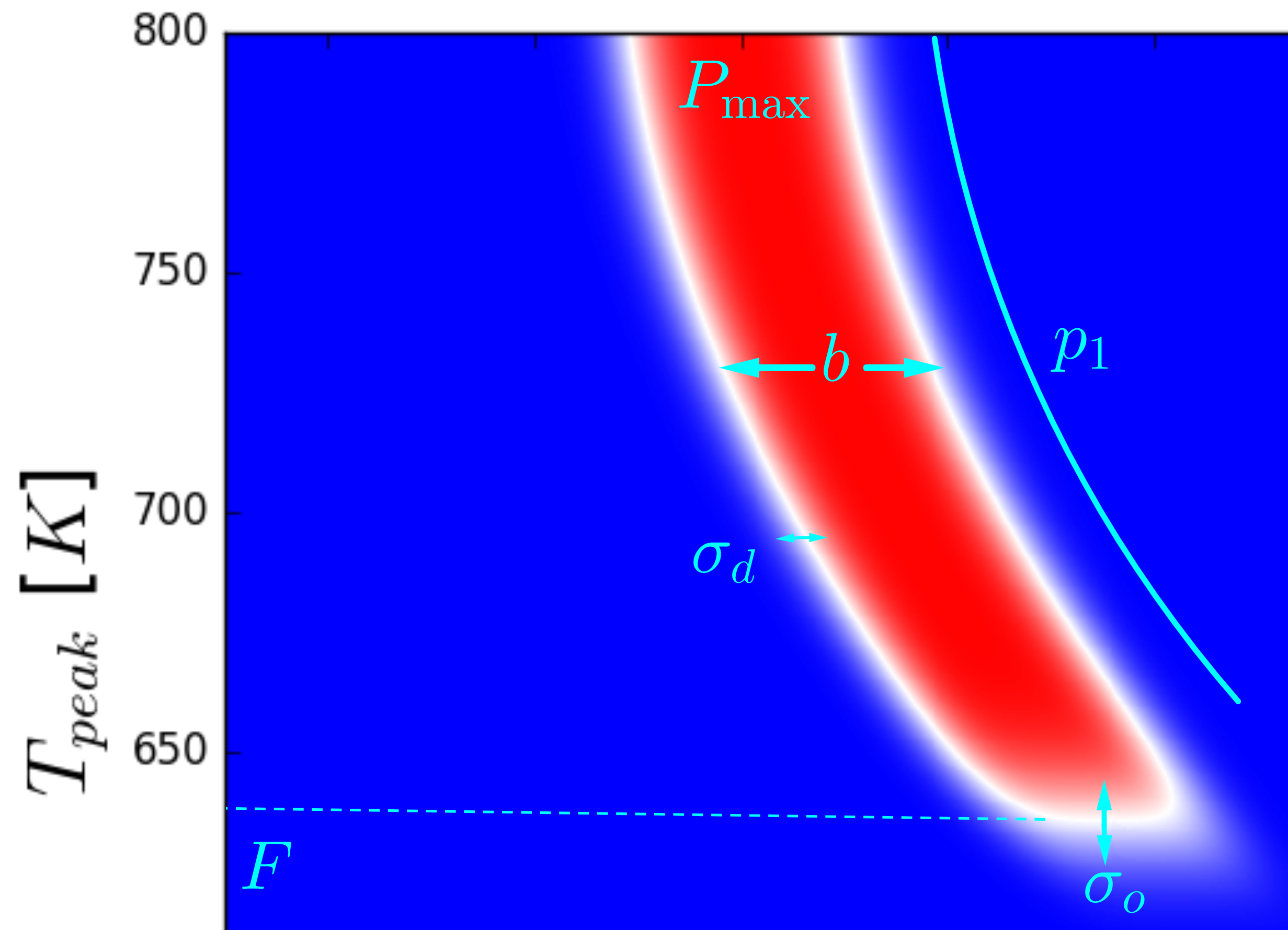}}
\caption{Detailed impact of the different model parameters in Sec. \ref{sec::modellparameters} visualized via an enlarged view of Fig. \ref{fig::progress} (c).}
\label{fig::parameterplot}
\end{figure}

We now define the model function $P$ in three steps with two help functions $h_1$ and $h_2$. Note that those functions depend on the upper parameters. First the down-track- and off-track-jitter is modeled as the slope of the probability function graph along a cut through the phase diagram for fixed $d$ and $T_\text{peak}$, respectively. As in Ref.~\cite[eq. (4)]{b2} we use the Gaussian cumulative distribution function
\begin{align}
\Phi (x,\mu,\sigma) = \frac{1}{2} \Bigg(1+\text{erf}\left(\frac{x-\mu}{\sqrt{2} \sigma}\right)\Bigg)
\end{align}
with
\begin{align}
\lim_{x\rightarrow -\infty} \Phi(x) = 0, \lim_{x\rightarrow +\infty} \Phi(x) = 1, \Phi(\mu) = \frac{1}{2}
\end{align}
and $\sigma$ determining the slope of $\Phi$. We write

\begin{align}
h_1(d,T) := \sqrt{P_\text{max}}\cdot \Phi(d,0,\sigma_d) \cdot \Phi(T,F,\sigma_o)
\end{align}

and receive a function as in Fig. \ref{fig::progress} (a) that models the down-track-jitter parameter $\sigma_d$ via the vertical and the off-track-jitter parameter $\sigma_o$ via the horizontal contour sharpness. The variable $\sqrt{P_\text{max}}$ determines the maximum function value, i.e. the formal limit
\begin{align}
\lim_{d\rightarrow \infty,T} h_1(d,T) = \sqrt{P_\text{max}}.
\end{align}
 When a bit is written, the switching probability should again decrease after a certain writing distance in down-track direction, therefore we additionally multiply a mirrored and shifted function $h_1$ in the form
 \begin{align}
h_2(d,T) := h_1(d,T) \cdot h_1(b-d,T) 
\end{align}

and receive a function graph as in Fig. \ref{fig::progress} (b). The maximum function value is $P_\text{max}$ but note, that this only holds in the limit for $b,T\rightarrow \infty$, so $P_\text{max}$ might never be actually reached. Finally we receive the complete model via transformation of the bit into a parabolic shape via

 \begin{align}
P(d,T) := h_2\Big(d-\big(p_1 (T-p_2)^2-p_3\big), T \Big)
\end{align}

 to get the C-like curvature as in Fig. \ref{fig::progress} (c), which is usually observed (see Ref.~\cite[Fig.2]{b7}). The impact of the model parameters defined in Sec. \ref{sec::modellparameters} on the shape of the model function can be visually observed in Fig. \ref{fig::parameterplot}. 
 
 \subsection{Reference system and variation of the parameters}\label{sec::parameters}
We aim to investigate the influence of the described switching probability phase plots on the resulting SNR. Of course, it is desirable to start all variations from a realistic basic parameter set. To compute a reference phase plot, we use the LLB model as in Ref.~\cite{b1} with the material parameters that can be seen in Tab.~\ref{tab::parameters}, a grain height of $8$~nm, an applied external magnetic field of $0.8$~T tilted with an angle of $22^\circ$ with respect to the $z$-direction and a duration of $0.67$~ns. Further a moving Gaussian heat pulse with a velocity of $v=15$~m/s and a full width at half maximum (FWHM) of $60$~nm, that leads to a maximum thermal gradient of $11$~K/nm, is assumed.

\begin{table}[b]
\begin{center}
\begin{tabular}{cc}
\hline\\
$K_1$ [J/$\text{m}^3$] & $6.6\cdot 10^6$\\
$J_{k,l}$ [J/link] & $6.72\cdot 10^{-21}$\\
$\mu_s$ [$\mu_B$] & $1.6$ \\
$J_s$ [T] & $1.35$ \\
$a$ [nm] & $0.24$ \\
$\lambda$ & $0.02$ \\
\hline
\end{tabular}
\end{center}
\caption{Used material parameters in the LLB model}
\label{tab::parameters}
\end{table}

\begin{figure}[htbp]
\centerline{\includegraphics[width=0.58\textwidth]{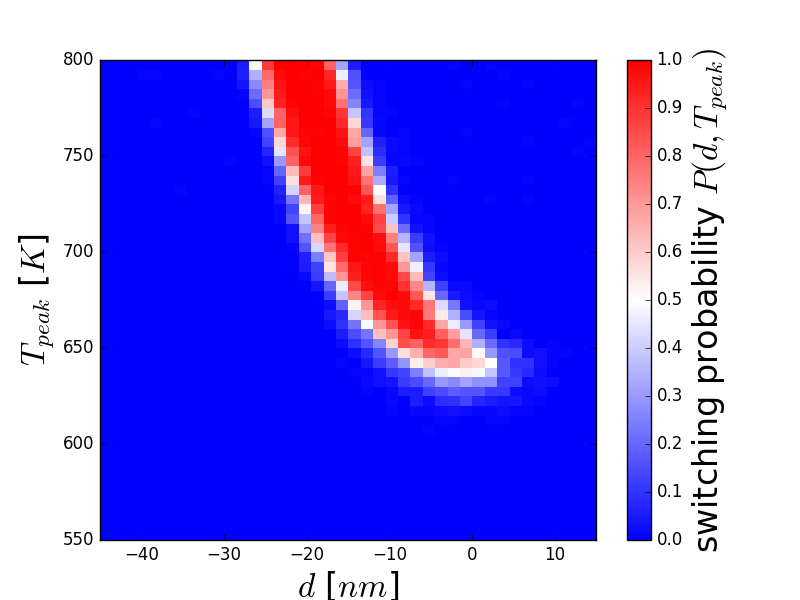}}
\caption{Simulated switching probability phase diagram with the LLB model for a grain diameter of $D=7$~nm and material parameters as shown in Tab.~\ref{tab::parameters}.}
\label{fig::original}
\end{figure}

 The resulting phase diagram for a grain diameter of $D=7$~nm is shown exemplarily in Fig. \ref{fig::original}. The reference parameters of the plots in Fig. \ref{fig::progress} are determined via a least square fit of the simulated diagram in Fig. \ref{fig::original}. For different grain sizes the eight resulting fitted parameters according to Sec. \ref{sec::modellparameters} are given in Tab.~\ref{tab::fit}.

\begin{table}[b]
\begin{center}
\begin{tabular}{|c|ccccc|}
\hline
&  \multicolumn{5}{|c|}{grain diameter}          \\
& $4$ nm & $5$ nm & $6$ nm & $7$ nm & $8$ nm \\
\hline
$\sigma_d$ [nm] & 2.51 & 2.03 & 1.74 & 1.50 & 1.37 \\
$\sigma_o$ [K] & 27.7 & 22.5 & 18.0 & 14.4 & 13.6\\
$P_\text{max}$ & 0.993 & 0.995 & 0.993 & 0.997 & 1.000 \\
$F$ [K] & 571 & 602 & 617 & 628 & 639 \\
$b$ [nm] & 10.2 & 10.2 & 10.3 & 10.2 & 10.1 \\
$p_1$ [$10^{-4}$ nm/$\text{K}^2$] & 3.28 & 3.88 & 4.33 & 4.89 & 5.16\\
$p_2$ [$\text{K}$] & 839 & 839 & 836 & 830 & 832\\
$p_3$ [nm] & 29.5 & 27.5 & 26.5 & 25.8 & 25.8\\
\hline
\end{tabular}
\end{center}
\caption{Reference parameters that are evaluated via least square fit of the simulated phase diagrams for grain diameters from $4$ to $8$ nm.}
\label{tab::fit}
\end{table}

\begin{figure}[htbp]
 \begin{center}
  \begin{subfigure}{0.35\textwidth}
\centerline{\includegraphics[width=1.00\textwidth]{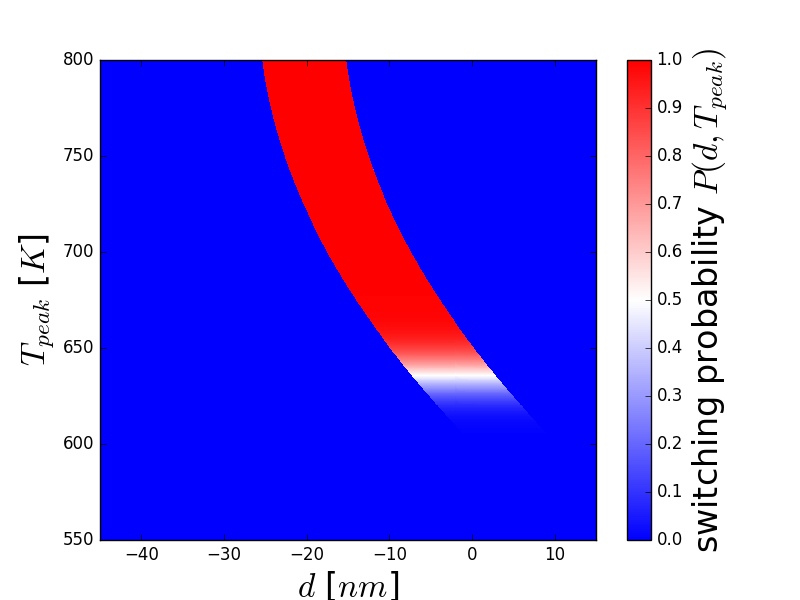}}
 \caption{}
 \end{subfigure}
 \begin{subfigure}{0.35\textwidth}
\centerline{\includegraphics[width=1.00\textwidth]{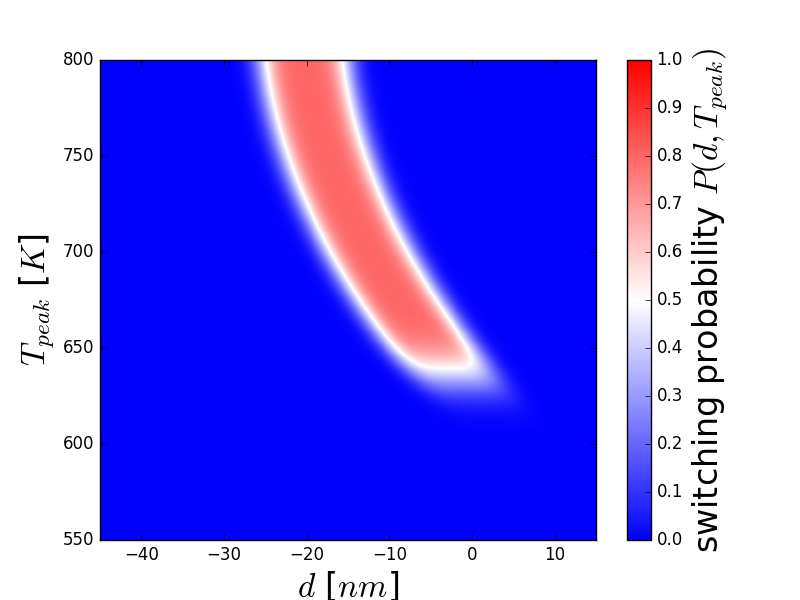}}
 \caption{}
 \end{subfigure}
   \begin{subfigure}{0.35\textwidth}
\centerline{\includegraphics[width=1.00\textwidth]{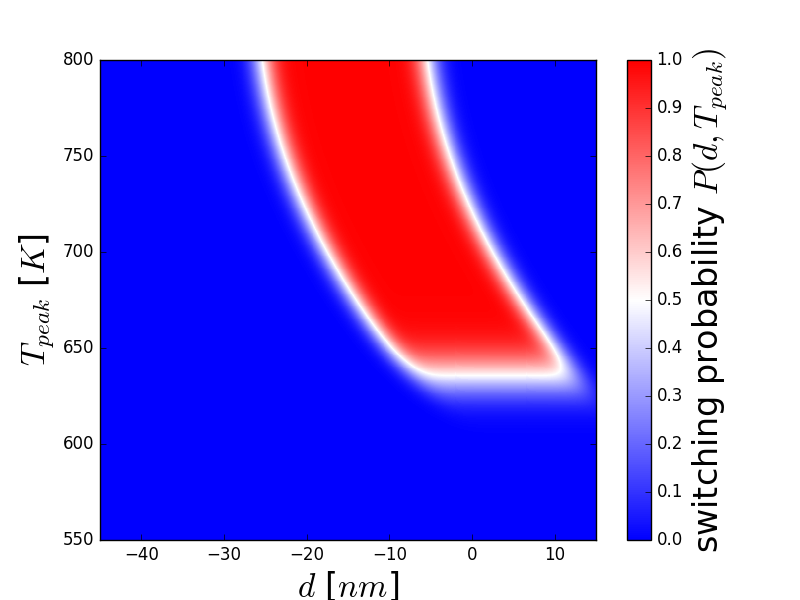}}
 \caption{}
 \end{subfigure}
    \begin{subfigure}{0.35\textwidth} 
\centerline{\includegraphics[width=1.00\textwidth]{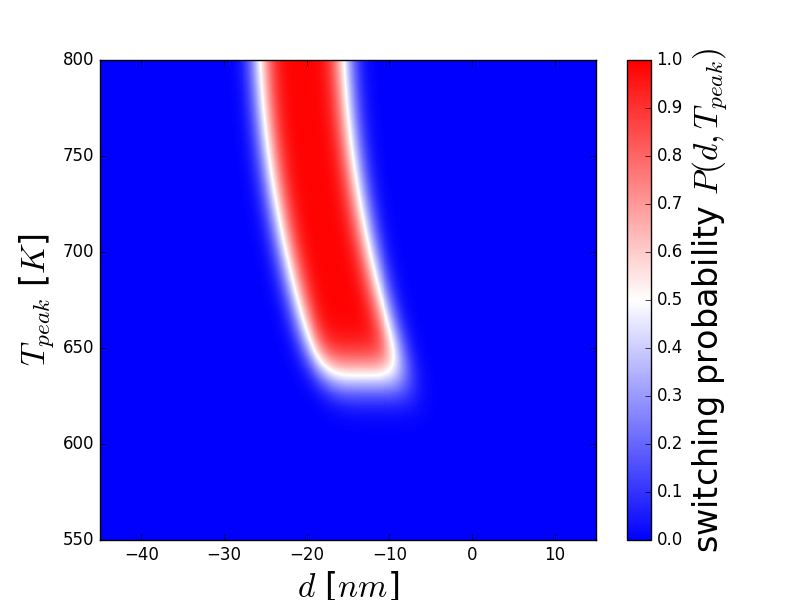}}
 \caption{}
 \end{subfigure}
\caption{In comparison to Fig. \ref{fig::progress} (c), each picture shows only one changed parameter value. The phase plot in (a) has a reduced down-track-jitter parameter $\sigma_d = 0.0001$ nm, in (b) a reduced $P_\text{max} = 0.8$, in (c) an extended bit-length of $b=20$ nm and in (d) a reduced curvature parameter $p_1$ by $60\%$.}
\label{fig::variations}
\end{center}
\end{figure}

In Sec.~\ref{sec::results} we investigate the influence of $\sigma_d, P_\text{max}, b$ and $p_1$ on the read-back SNR. The investigated parameter variations are taken in the ranges of Tab.~\ref{tab::variations} and in Fig. \ref{fig::variations} (a)-(d) their influence on the phase diagram is visualized.

\begin{table}
\begin{center}
\begin{tabular}{|c|c|c|}
\hline
parameter & min value & max value \\
\hline
$\sigma_d$ [nm] & 0.01 & 4.00 \\
$P_\text{max}$ & 0.64 & 1.00 \\
$b$ [nm] & 4.0 & 12.0 \\
curvature reduction [\%] & 0 & 100 \\
\hline
\end{tabular}
\end{center}
\caption{Range of variation for the model parameters.}
\label{tab::variations}
\end{table}

\section{Bit patterns on granular media}\label{sec::printing}
\subsection{Writing process}

\begin{figure}[htbp]
\centerline{\includegraphics[width=0.50\textwidth]{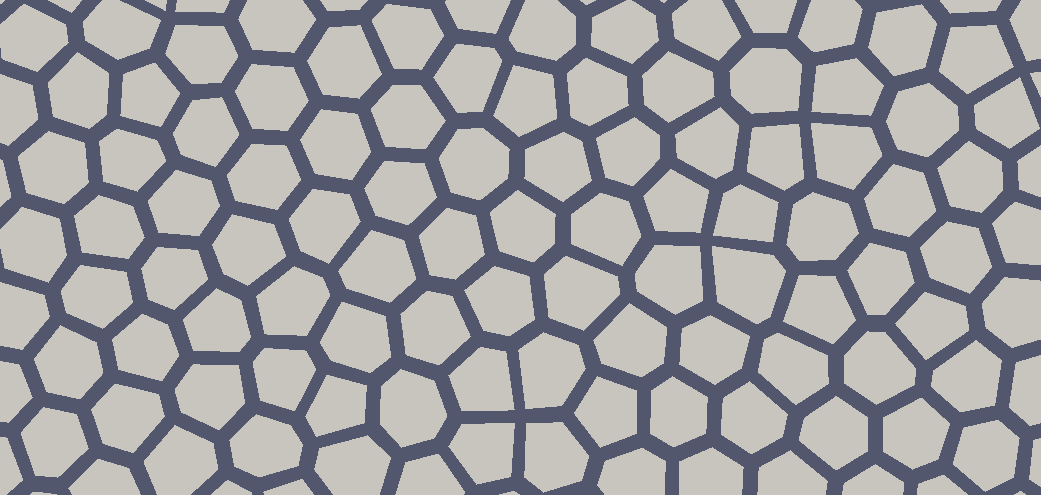}}
\caption{Visualization of the granular medium with approximately equally sized magnetic grains (here: $4$~nm diameter) surrounded by nonmagnetic material separating neighboring grains by $1$~nm.}
\label{fig::grains}
\end{figure}

\begin{figure}[htbp]
\centerline{\includegraphics[width=0.50\textwidth]{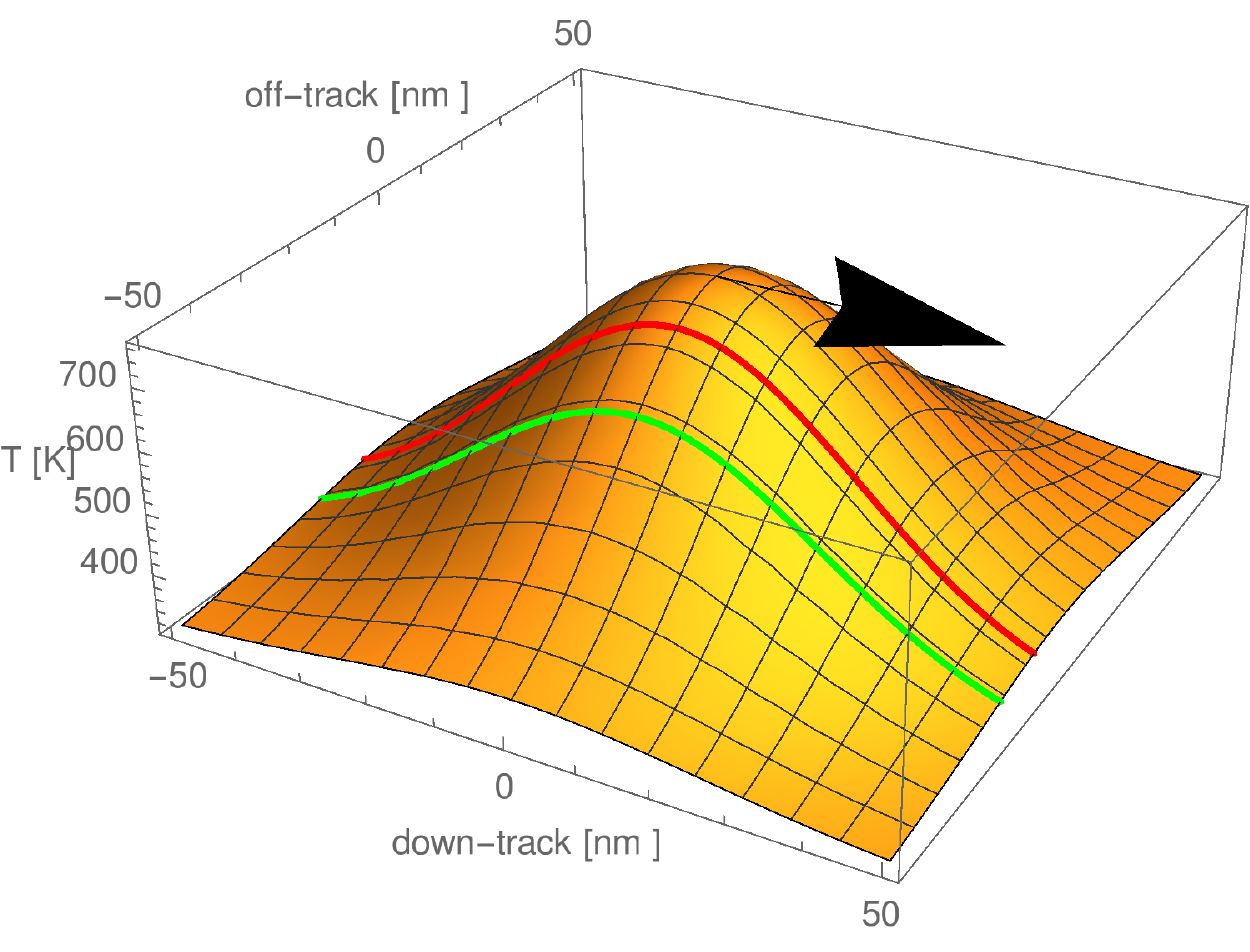}}
\caption{Visualization of the Gaussian heat pulse that moves across the granular medium in the direction of the arrow. Together with an applied magnetic field it performs the writing process. The red and green curve demonstrate that grains are exposed to different peak temperatures depending on their off-track position.}
\label{fig::gauss}
\end{figure}

\begin{figure}[htbp]
\centerline{\includegraphics[width=0.50\textwidth]{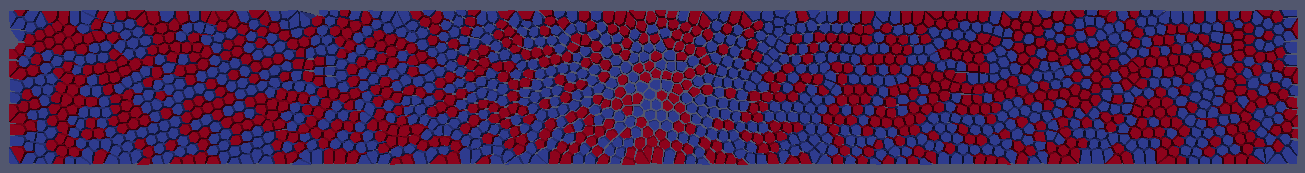}}
\centerline{\includegraphics[width=0.50\textwidth]{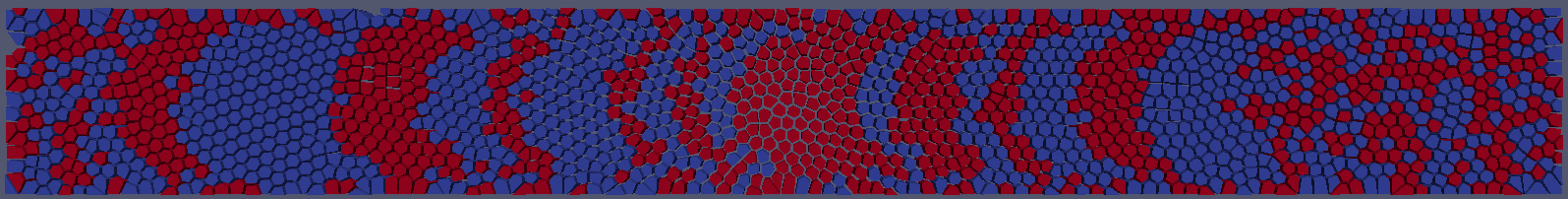}}
\caption{Top: Randomly initialized granular medium ($500\times 60$ nm and a thickness of $8$ nm) with grain diameter of $4$ nm and $1$ nm gap between neighboring grains. The colors distinguish magnetizations in up or down direction (blue: down, red: up). Bottom: Granular medium after the simulated writing process.}
\label{fig::writing}
\end{figure}

We aim to use a phase plot, which determines the switching probability of a single cylindrical grain, to write bit patterns on granular media. Each medium contains approximately equally sized magnetic grains with a diameter of $D = 4,5,6,7$ respectively $8$~nm. The diameter's standard deviation of about $0.31$~nm is neglected in the following writing process, so we assume that each grain within the granular medium has the same size. Nonmagnetic material separates neighboring grains by $B=1$~nm (see Fig. \ref{fig::grains}, visualized exemplarily for a diameter of $4$~nm). We assume a Gaussian heat pulse moving across the medium with a velocity of $v = 15$~m/s, a full width at half maximum (FWHM) of $60$~nm and a maximum temperature of $T_C + 60$~K, which is slightly dependent on the Curie temperature $T_C$ for different grain sizes of about $700$~K (see Fig. 5). The writing process is justified due to the assumption that every grain is approximated by a cylinder with a certain diameter that is subject to the Gaussian heat pulse and a trapezoidal magnetic field. The peak temperature of the heat pulse $T_{\text{peak}}$ depends on the off-track position $y$ of the grain (see Fig. \ref{fig::gauss}) via Eq.~\ref{eq::peakofftrack}. We neglect a spatially varying temperature within a single grain and therefore assume, that the whole grain volume receives the same temperature within the entire grain. The magnetic field is aligned according to the written pseudo random bit series of Ref.~\cite{b3}: 
\begin{align*}
(\text{-}1\ \text{+}1\ \text{+}1)\ \text{-}1\ \text{-}1\ \text{-}1\ \text{-}1\ \text{-}1\ \text{+}1\ \text{+}1\ \text{+}1\ \text{-}1\ \text{-}1\ \text{+}1\ \text{-}1\ \text{-}1\ \text{-}1\ \text{+}1\ \text{-}1\\
 \text{+}1\ \text{-}1\ \text{+}1\ \text{+}1\ \text{+}1\ \text{+}1\ \text{-}1\ \text{+}1\ \text{+}1\ \text{-}1\ \text{+}1\ \text{-}1\ \text{-}1\ \text{+}1\ \text{+}1\ (\text{-}1\ \text{-}1\ \text{-}1),
\end{align*}
 where $\text{-}1$ represents a down and $\text{+}1$ an up bit (i.e. field in negative, respectively positive direction). A sequence of two or more successive down respectively up bits is summed up to one writing process with multiple bit length. The stray-field is not taken into account directly and thus different grains do not influence the switching probability of each other. Its impact is only considered via an added variation to $T_C$ as discussed in Ref.~\cite{b12}. The writing is done by mapping the switching probabilities of the corresponding phase plot to the grains of the granular medium according to their positions. Therefore a uniformly distributed pseudo random number $r\in [0,1)$ is created. If the probability $p$ fulfills $p>r$, the magnetization of the grain is changed according to the direction of the applied field. Otherwise it remains unchanged in the previous direction. The $2\times 3$ bits in the brackets are used for padding and the remaining $31$ bits in-between represent the desired bit series. The result of the whole process is shown in Fig. \ref{fig::writing}.

\subsection{Reading process}\label{sub::reading}
\begin{figure}[htbp]
\begin{subfigure}{0.50\textwidth} 
\centerline{\includegraphics[width=1.00\textwidth]{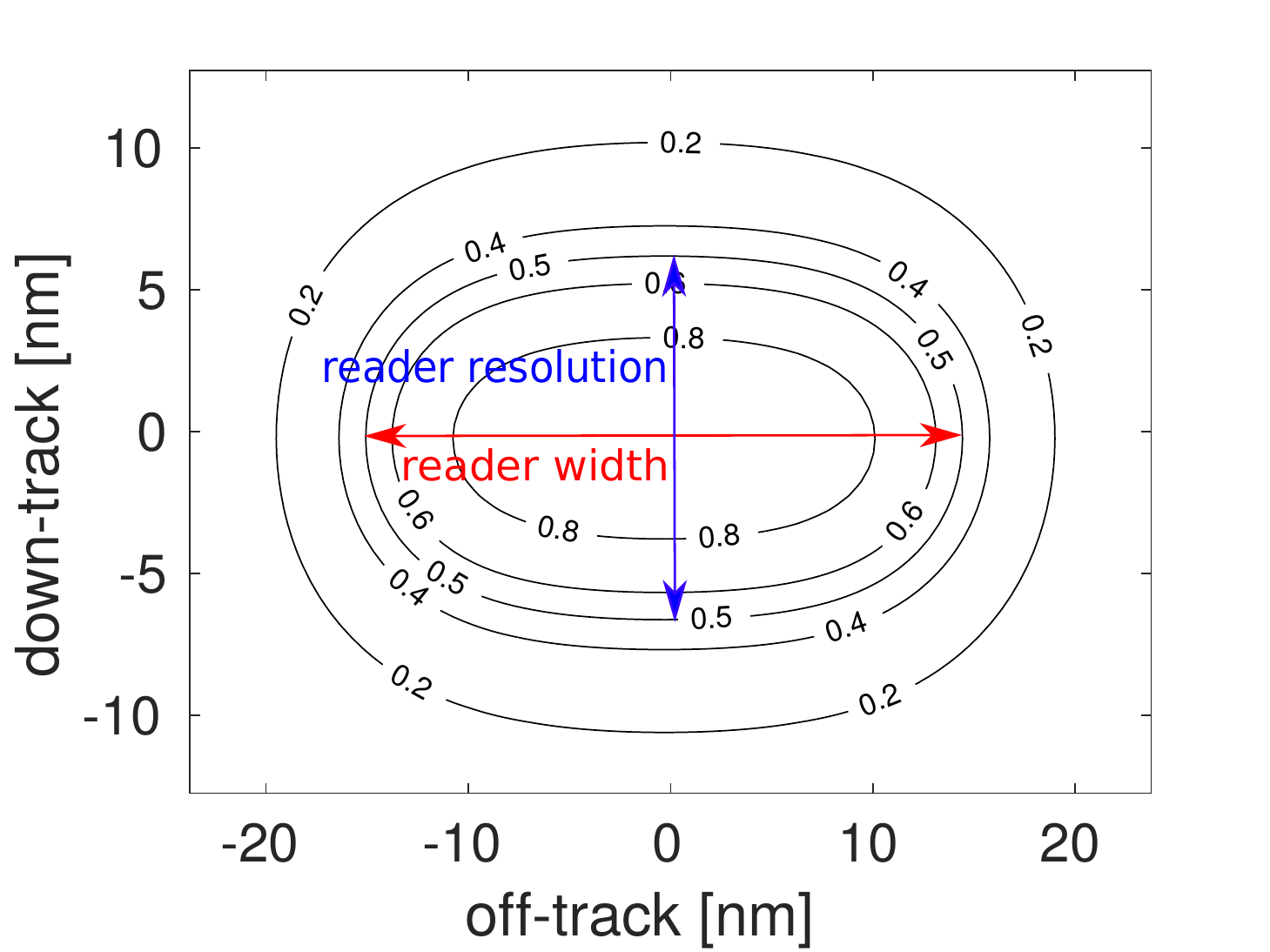}}
\caption{}
\end{subfigure}
\begin{subfigure}{0.50\textwidth} 
\centerline{\includegraphics[width=1.00\textwidth]{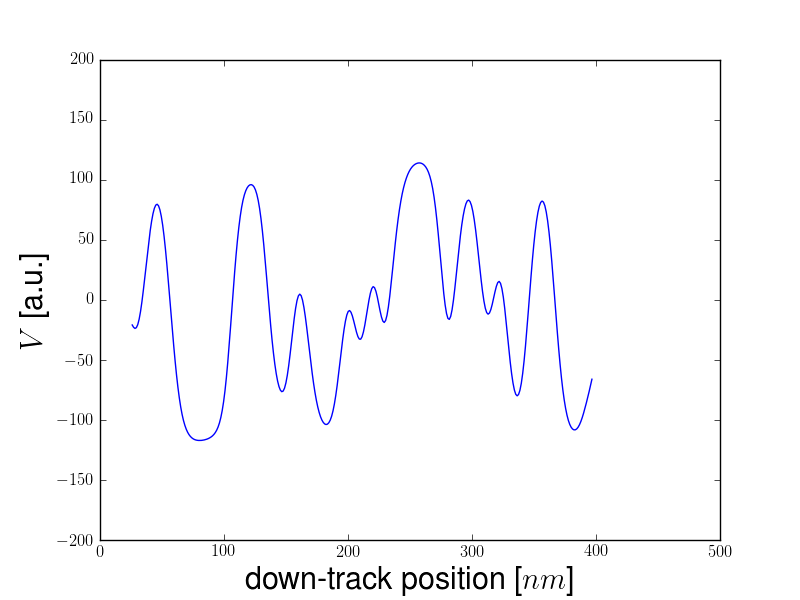}}
\caption{}
\end{subfigure}
\caption{(a): Contour plot of the sensitivity function $H_z$ with a data point resolution of $0.5$ nm in down- and off-track-direction. The reader width (width between $50\%$ amplitude points in off-track direction) is $30.13$ nm and the reader resolution (width between $50\%$ amplitude points in down-track direction) is $13.26$ nm. (b) Example for a read-back curve determined by the reader with the sensitivity function in (a) across the granular medium in Fig. \ref{fig::writing}, bottom.}
\label{fig::sensitivity}
\end{figure}

The reader module is defined via its sensitivity function as discussed in Ref.~\cite{b8}. It is illustrated in Fig. \ref{fig::sensitivity} (a). The voltage $V$ of the reader is given by the integral in Ref.~\cite[Eq.~(1)]{b8}
\begin{align}
V = c_1 \cdot \int \overrightarrow{H}\cdot \overrightarrow{M}\ \text{d}V_m
\end{align}
with the reader's sensitivity function $\overrightarrow{H}$, the media's magnetization $\overrightarrow{M}$ and a reader dependent constant $c_1$. Since the magnetization is assumed to have negligible dependence on the $z$-direction and grains have strong axial anisotropy, it degenerates to an area integral in the form
\begin{align}
V = \widetilde{c}_1 \cdot \int H_z M_z \ \text{d}A_m
\end{align}
with a constant $\widetilde{c}_1$, which does not affect the SNR value. This integral is computed via a discrete sum over the data points of the sensitivity function, multiplied with $-1, +1$ or $0$ depending whether the data point is within a grain with magnetization in down- or up-direction or a gap. Moving the sensitivity function in $0.5$ nm steps across the medium gives the detected read-back signal of the whole bit pattern as shown in Fig. \ref{fig::sensitivity} (b). 

\subsection{SNR calculation}\label{sub::snr}
\begin{figure}[htbp]
\centerline{\includegraphics[width=0.50\textwidth]{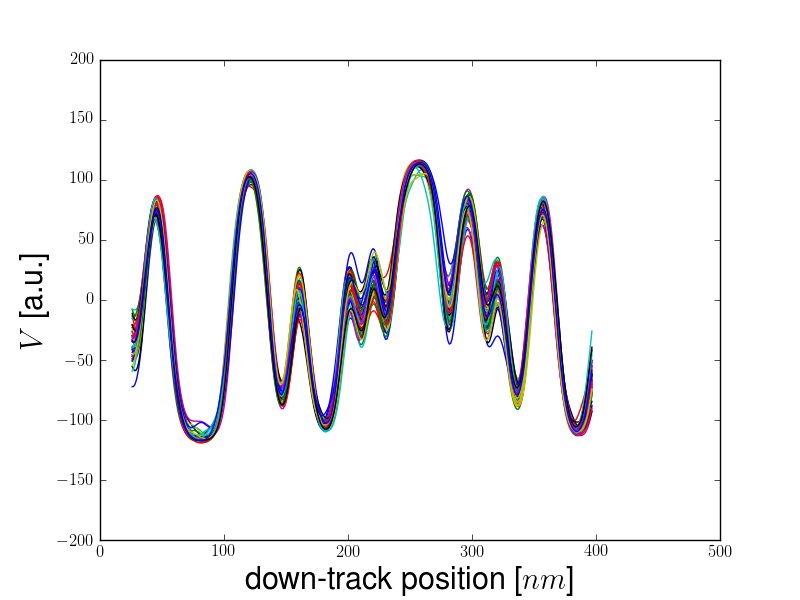}}
\caption{The read-back signal of $50$ bit patterns written with the same model phase plot, but on different granular media. The shape of the curves is used for the SNR calculation of the signal.}
\label{fig::50curves}
\end{figure}

By repeating the writing and reading process on $50$ different randomly initialized granular media, we are able to calculate the SNR with the methods of Ref.~\cite{b3}. 
The readback signal of the same bit sequence on 50 different media is shown in Fig.~\ref{fig::50curves}.

\section{Results and Discussion}\label{sec::results}

\subsection{SNR curves}\label{sub::snr}

\begin{figure}[htbp]
\centerline{\includegraphics[width=0.40\textwidth]{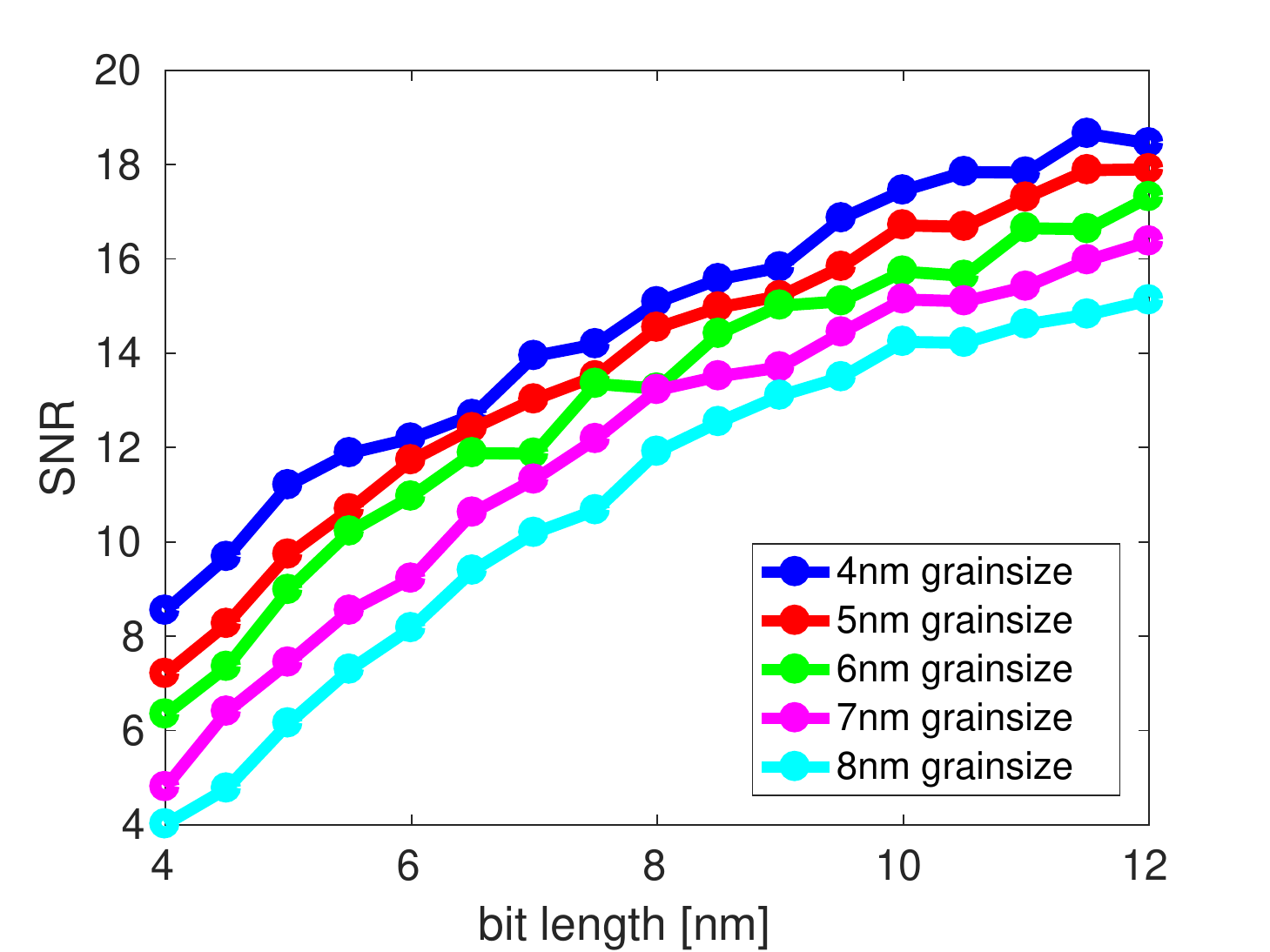}}
\centerline{\includegraphics[width=0.40\textwidth]{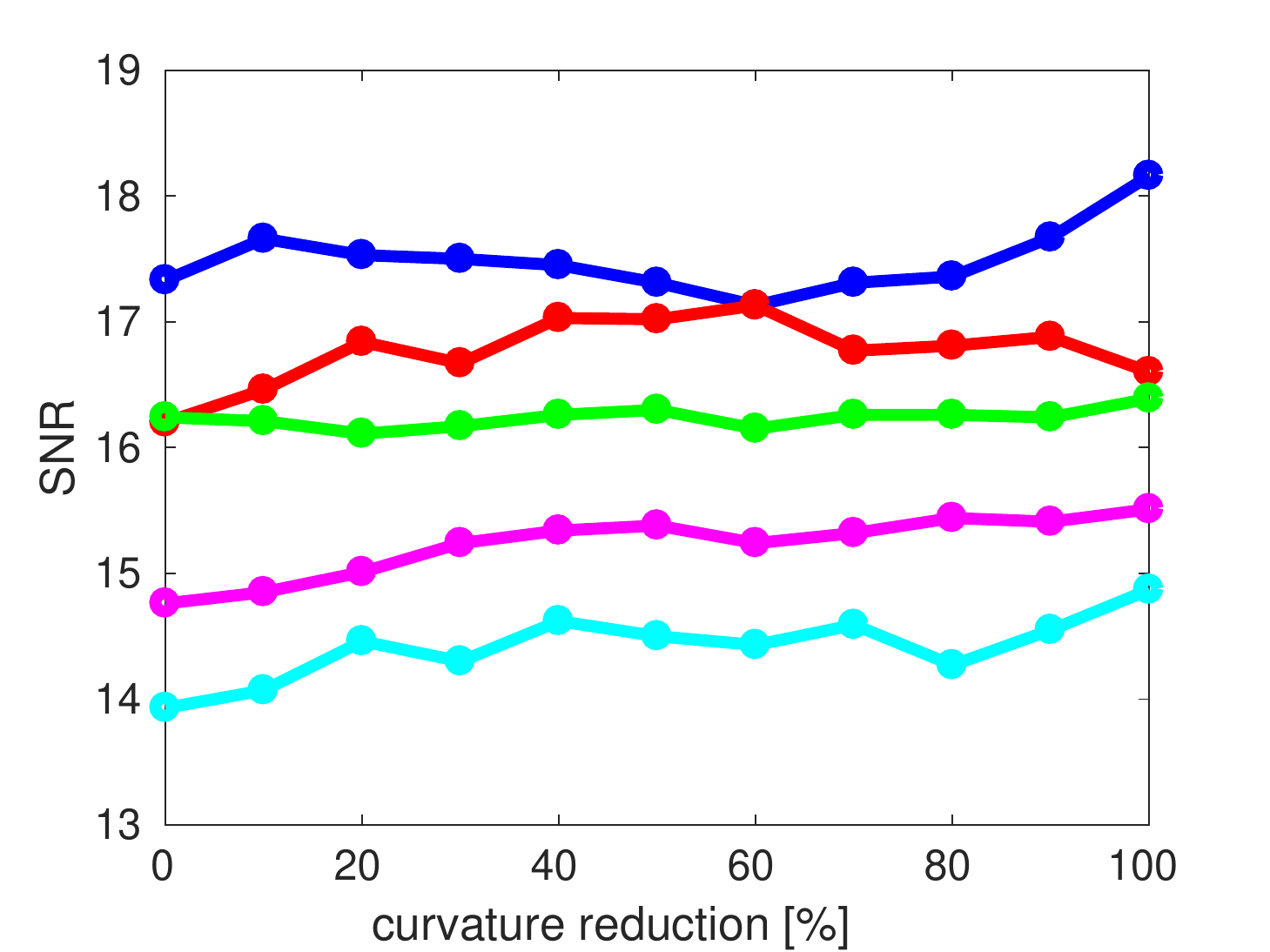}}
\centerline{\includegraphics[width=0.40\textwidth]{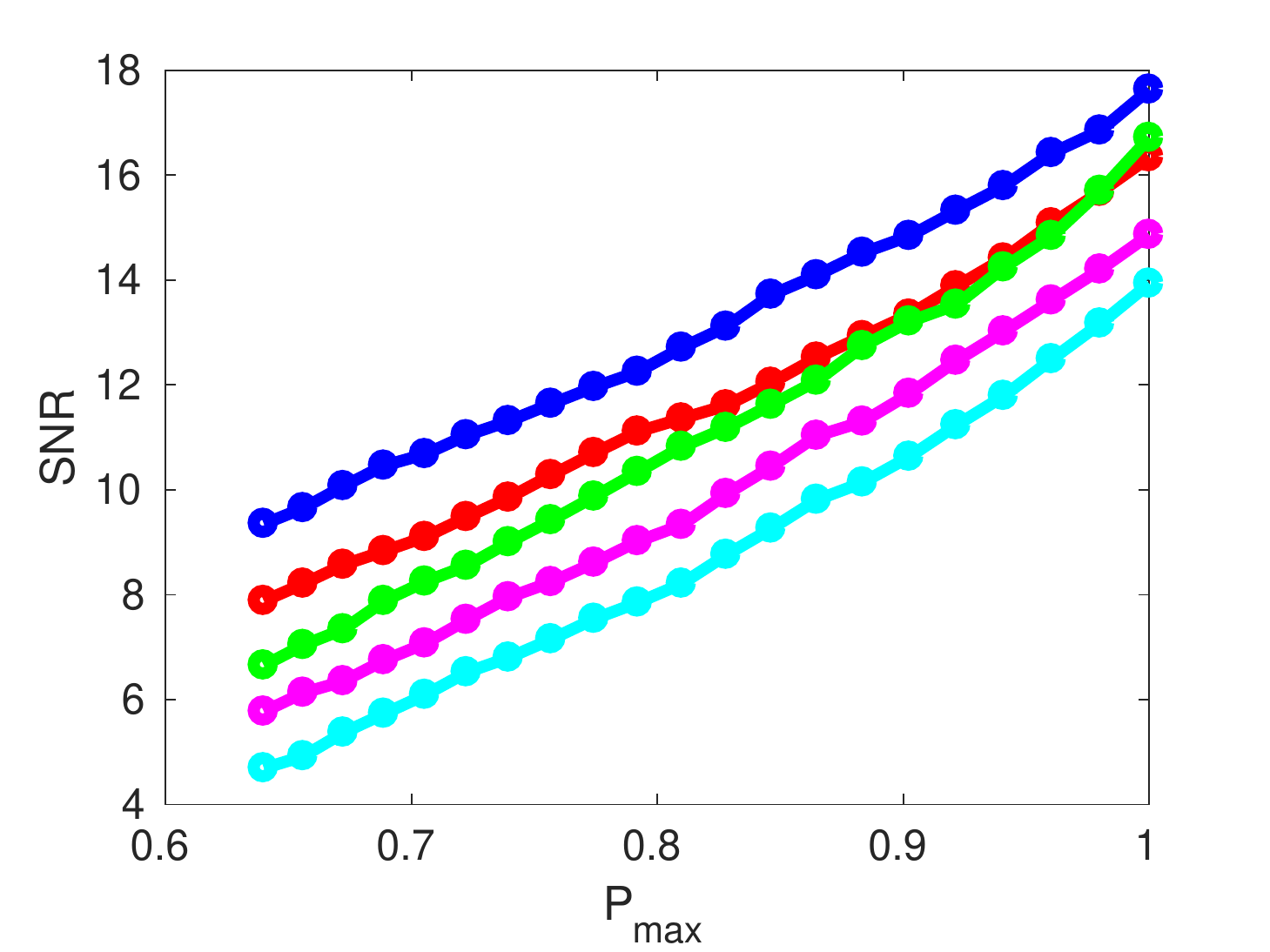}}
\centerline{\includegraphics[width=0.40\textwidth]{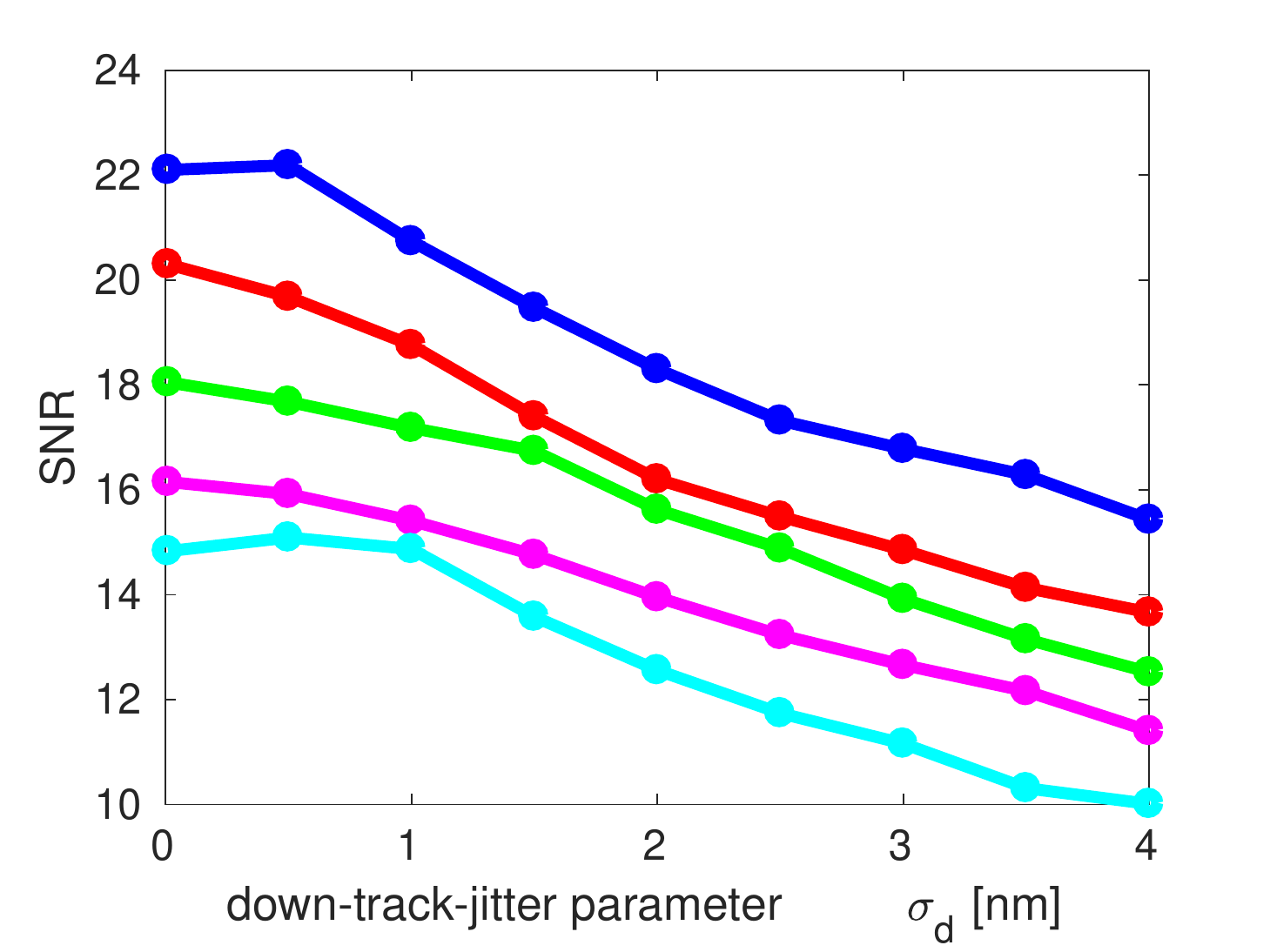}}
\caption{SNR as a function of: (a) the bit length (with scaled reader), (b) curvature reduction, (c) $P_{\text{max}}$ and (d) $\sigma_{d}$.}
\label{fig::results}
\end{figure}

For various parameter variations, the resulting SNR curves are plotted in Fig. \ref{fig::results}. For selected values of the parameters, Figs. \ref{fig::grains1} - \ref{fig::grains2} visualize the corresponding bit pattern. The impact of the varied parameters on the SNR can clearly be observed. The curves describing the dependence on the bit length, $P_\text{max}$ and the down-track-jitter $\sigma_d$ demonstrate an increase of $\sim\!\!10$~dB for bit lengths between $4$ and $12$~nm, an increase of $\sim\!\!9$~dB for maximum switching probabilities between $0.64$ and $1.00$ and a decrease of $\sim\!\!5$~dB for down-track-jitter parameters between $0$ and $4$~nm.
Note that for small bit lengths the reader needs a better resolution (see Fig. \ref{fig::sensitivity} (a)) to achieve a suitable read-back signal, therefore we scale the reader resolution in down-track direction $R$ according to the bit length $b$ in the form
\begin{align}\label{eq::scaling}
R = R_0 \cdot \frac{b}{b_0},
\end{align}
where $R_0 = 13.26$~nm denotes the initial reader resolution (see Fig. \ref{fig::sensitivity} (a)) and $b_0 = 10.2$~nm is the mean initial bit length according to the phase diagram in Tab.\ref{tab::fit}) . The results in Fig. \ref{fig::results}, top show a sharp decrease of the SNR for low bit lengths, so the possible reachable linear density is limited. $P_\text{max}$ and $\sigma_d$ are clearly two parameters with significant impact on the SNR, so it is recommendable to consider those values in terms of material optimization as in Ref.~\cite{b2}.
Furthermore the variation of two parameters simultaneously lead to SNR contours as in Fig.~\ref{fig::transition_pmax}. 

\begin{figure}[htbp]
\begin{subfigure}{0.50\textwidth} 
\centerline{\includegraphics[width=1.0\textwidth]{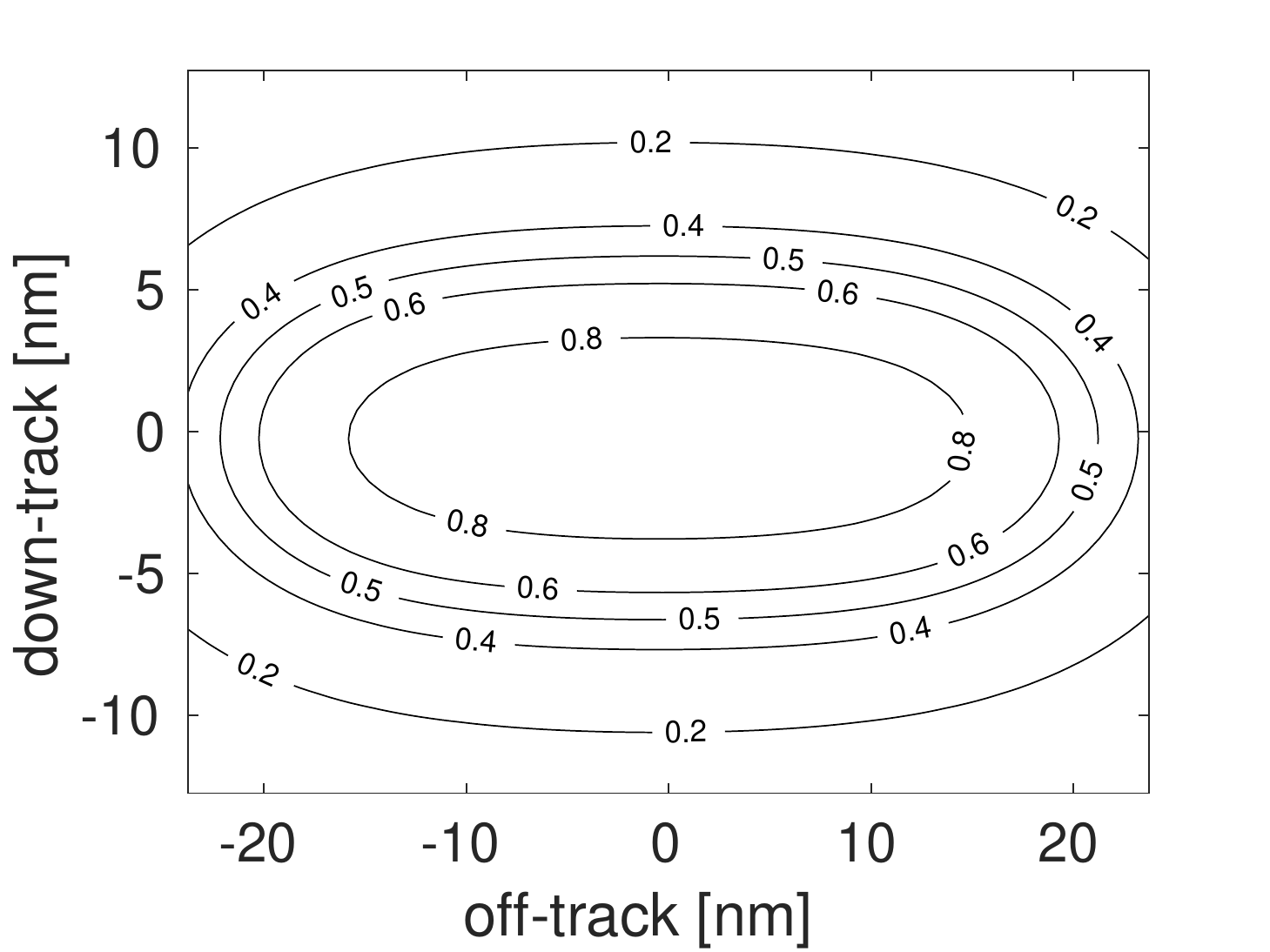}}
\caption{}
\end{subfigure}
\begin{subfigure}{0.50\textwidth} 
\centerline{\includegraphics[width=1.0\textwidth]{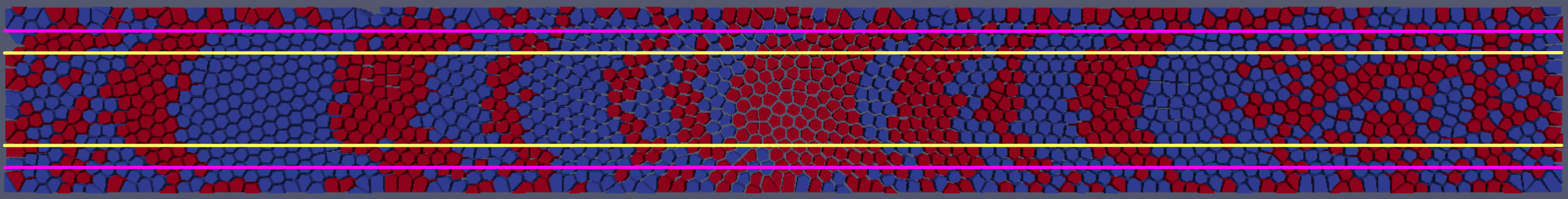}}
\caption{}
\end{subfigure}
\begin{subfigure}{0.50\textwidth} 
\centerline{\includegraphics[width=1.0\textwidth]{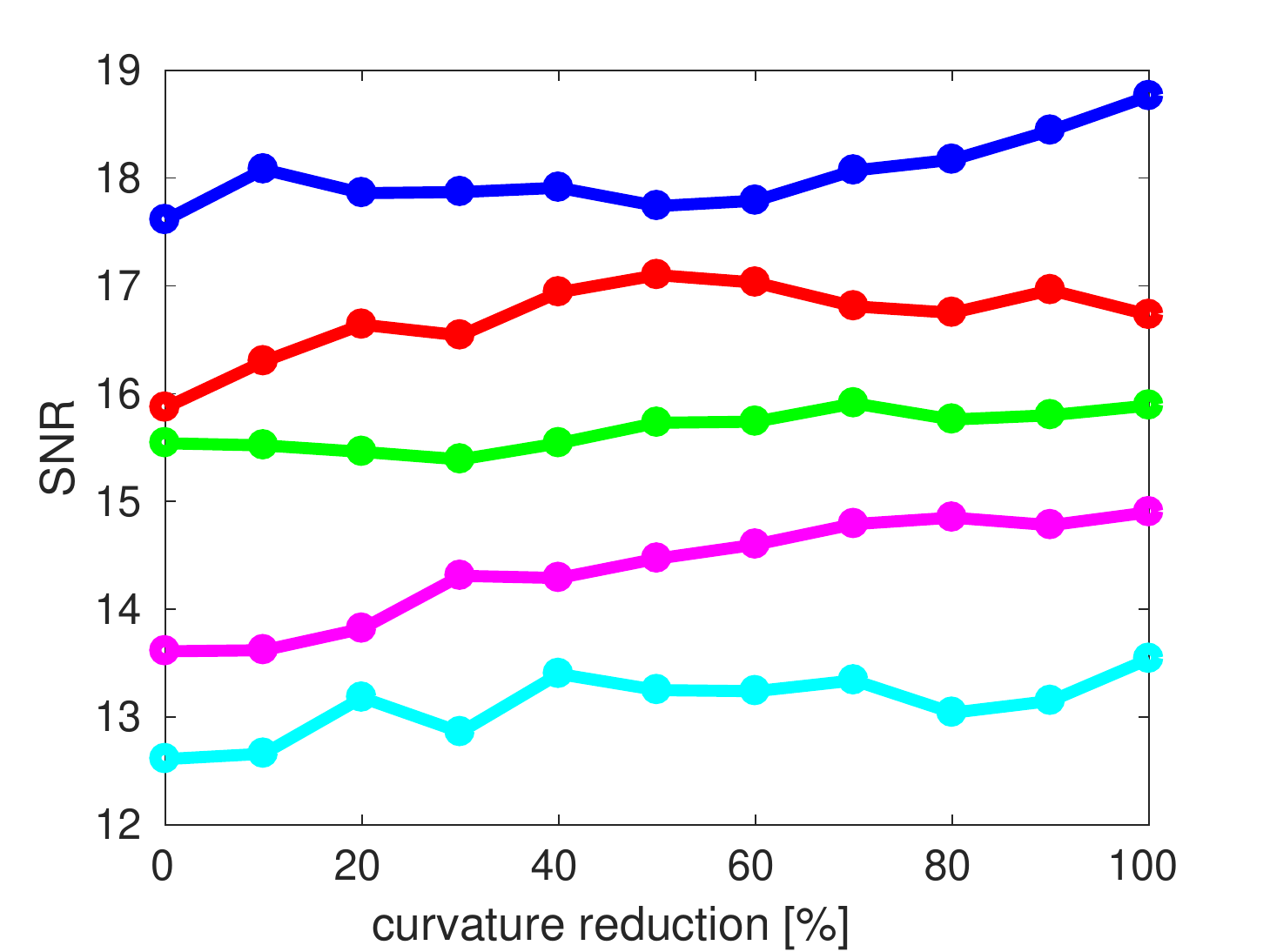}}
\caption{}
\end{subfigure}
\caption{(a): Contour plot of the scaled sensitivity function $H_z$ to a reader width of $44.34$ nm. (b): The initial (yellow) and new (pink) reader track edges. (c): The corresponding SNR values as a function of the curvature reduction computed with the scaled sensitivity function. }
\label{fig::scaledreader}
\end{figure}

\begin{figure}[htbp]
\centerline{\includegraphics[width=0.50\textwidth]{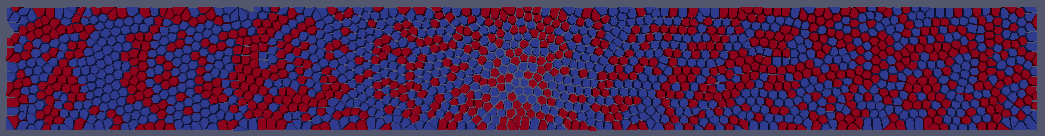}}
\centerline{\includegraphics[width=0.50\textwidth]{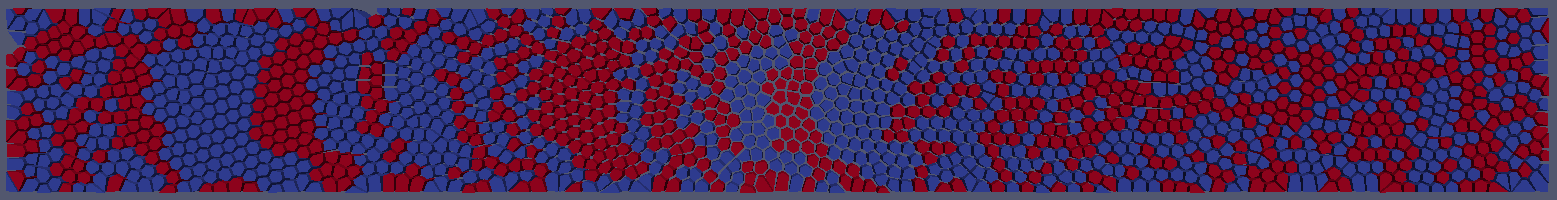}}
\centerline{\includegraphics[width=0.50\textwidth]{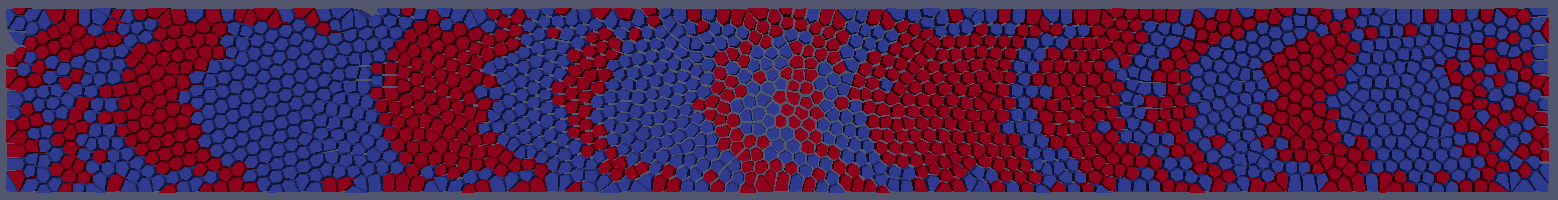}}
\caption{Bit pattern for bit lengths of $4, 7$ and $12$ nm (top to bottom).}
\label{fig::grains1}
\end{figure}

Within the chosen reference system, the bit curvature does not influence the SNR strongly ($0.5$ to $1$~dB). To ensure that the SNR improvement is not limited by the reader width we scale the reader width to the track width (which can be calculated with the half maximum temperature $F$ in Tab.~\ref{tab::fit} and the shape of the Gaussian heat pulse in Fig. \ref{fig::gauss} and has a mean value of $44.34$ nm) and calculate the read-back signal again (see Fig. \ref{fig::scaledreader} (a) and (b)). In this way we aim to receive more improvement due to the curvature reduction effect on the edges of the writing track. However, we can see in Fig. \ref{fig::scaledreader} (c) that the SNR slope remains almost the same.

\begin{figure}[htbp]
\centerline{\includegraphics[width=0.50\textwidth]{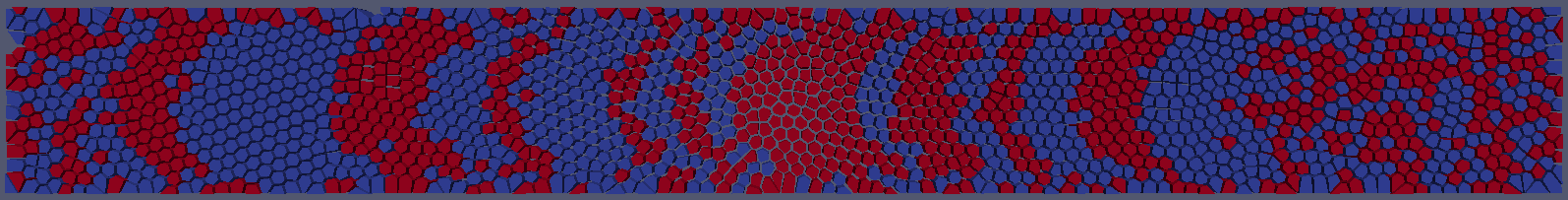}}
\centerline{\includegraphics[width=0.50\textwidth]{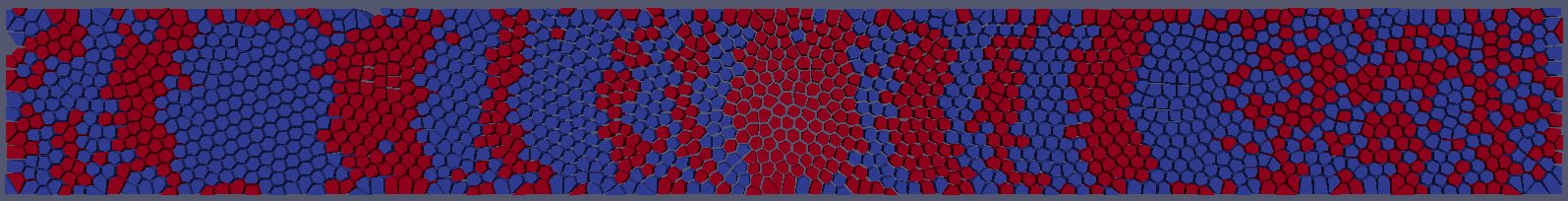}}
\centerline{\includegraphics[width=0.50\textwidth]{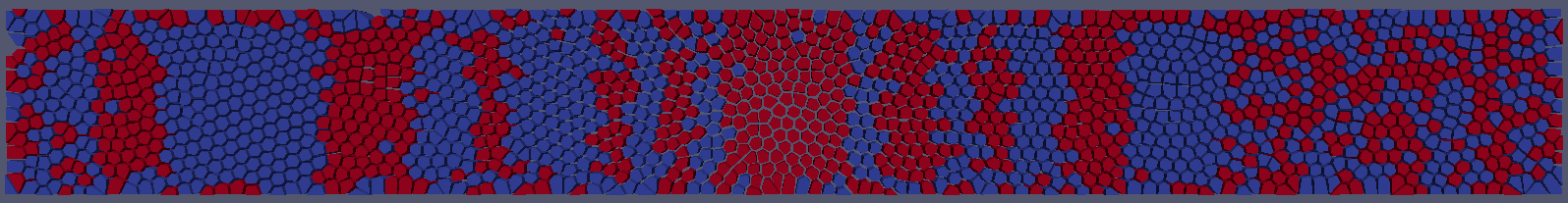}}
\caption{Bit pattern for curvature reductions of $0, 50$ and $100$ \% (top to bottom).}
\end{figure}

\begin{figure}[htbp]
\centerline{\includegraphics[width=0.50\textwidth]{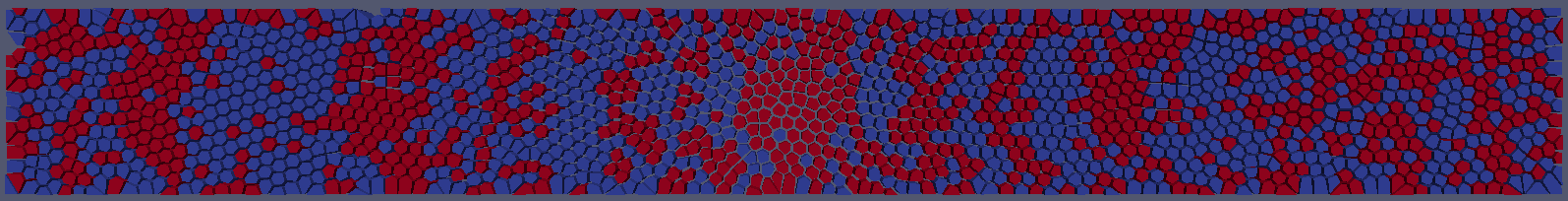}}
\centerline{\includegraphics[width=0.50\textwidth]{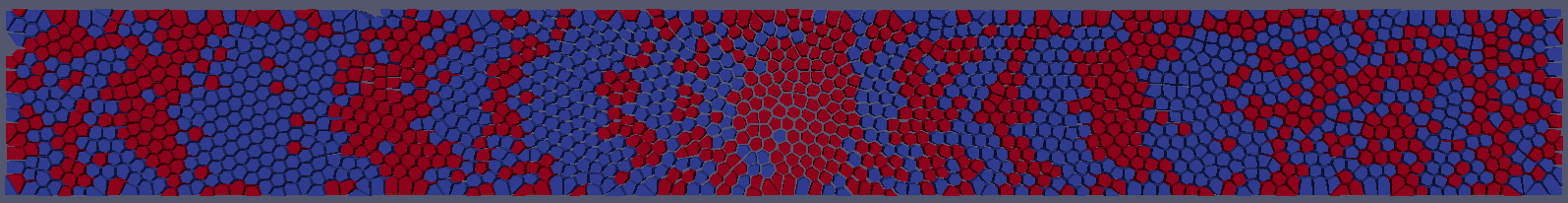}}
\centerline{\includegraphics[width=0.50\textwidth]{pmax_1p00.png}}
\caption{Bit pattern for $P_\text{max} = 0.64, 0.81$ and $1.00$ (top to bottom).}
\end{figure}

\begin{figure}[htbp]
\centerline{\includegraphics[width=0.50\textwidth]{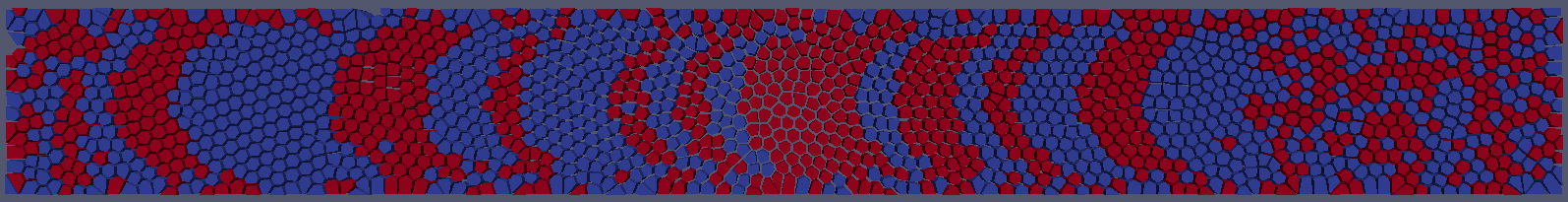}}
\centerline{\includegraphics[width=0.50\textwidth]{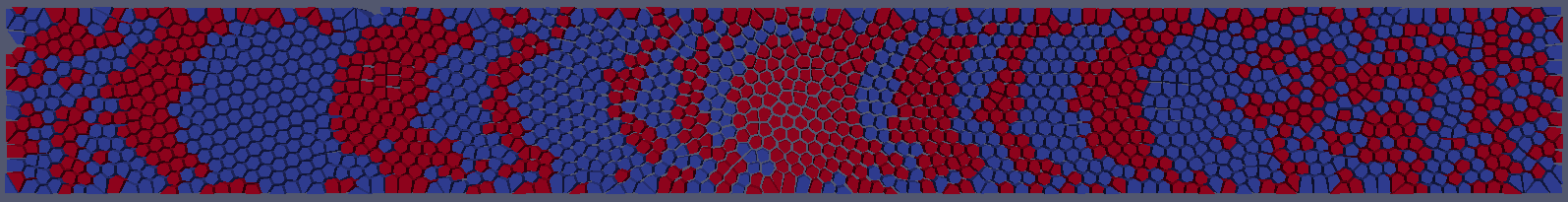}}
\centerline{\includegraphics[width=0.50\textwidth]{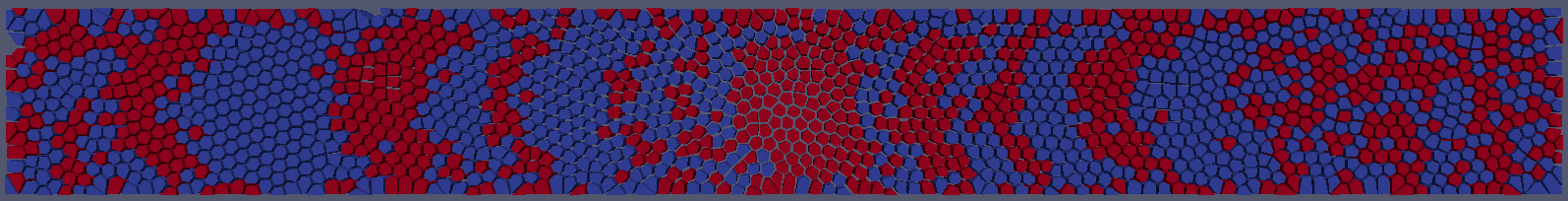}}
\caption{Bit pattern for $\sigma_d = 0.01, 2.00$ and $4.00$ nm (top to bottom).}
\label{fig::grains2}
\end{figure}

\begin{figure}[htbp]
\centerline{\includegraphics[width=0.50\textwidth]{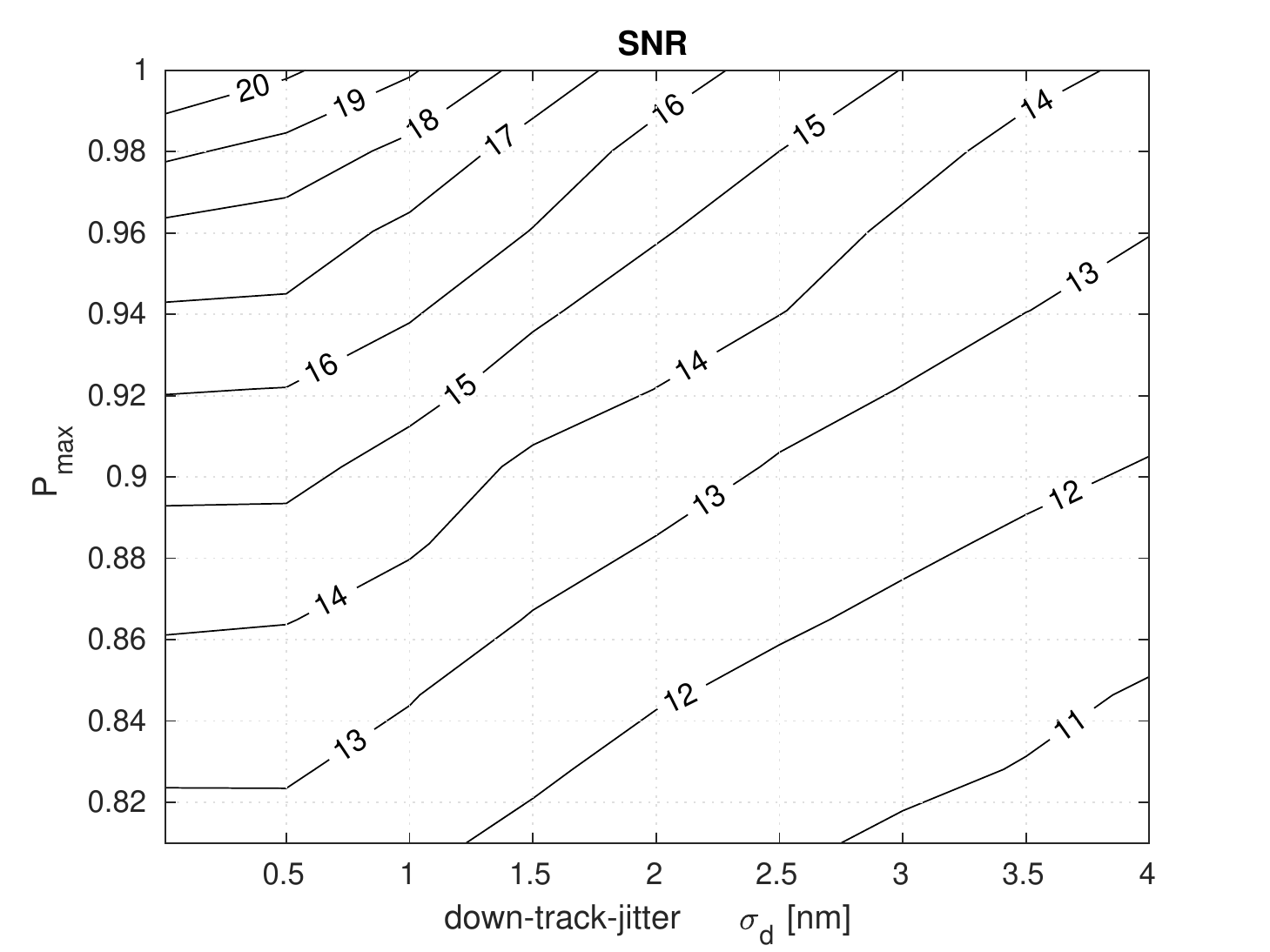}}
\caption{SNR values for simultaneous variation of $\sigma_d$ and $P_{\text{max}}$ for $5$ nm grain diameter.}
\label{fig::transition_pmax}
\end{figure}

\subsection{Comparison with theory}\label{sub::theory}
In Ref.~\cite[Eq.~(2.4)]{b5} the SNR-dependence on the media magnetic transition parameter $a$, grain size $S$ ($S=D+B$, i.e. sum of grain diameter $D$ and nonmagnetic boundary $B$, see Fig. \ref{fig::grains}), bit length $b$, read-back pulse width $T_{50}$ and reader width $W$ is given by
\begin{align}\label{eq::prop}
\text{SNR} \propto   \left(\frac{b}{a}\right)^2 \cdot \frac{T_{50}}{b} \cdot \frac{W}{S}.
\end{align}
As in Ref.~\cite[Eq.~(4)]{b9}, the media magnetic transition parameter $a$ can be separated into two independent parts:
\begin{itemize}
\item the down-track-jitter parameter $\sigma_d$, that originates from the probability distribution in the phase diagram
\item the grain-jitter parameter $\sigma_g$, that comes from the grain distribution in the granular medium
\end{itemize}
The stochastic independence ensures that the total jitter of these two parts can be written as 
\begin{align}
a = \sqrt{\sigma_d^2 + \sigma_g^2}.
\end{align}
The grain-jitter parameter $\sigma_g$ usually depends on the mean and the variation of the grain sizes. In our simulations, we assume the distribution to be very sharp, i.e. we neglect the dependence on the variations and assume all grains to have approximately the same size $S$. In this case $\sigma_g$ only depends on $S$ and a very simplified model with square grains in Ref.~\cite{b9} shows
\begin{align}\label{eq::model}
\sigma_g = \frac{S}{\sqrt{12}}.
\end{align}

Since the size of the nonmagnetic grain boundary is constant $B=1$~nm in our media (see Fig. \ref{fig::grains}), the grain size $S=D+B$ only depends on the grain diameter $D$. Furthermore, the proportionality in Eq. \eqref{eq::prop} may only hold, if the read-back pulse width $T_{50}$ is chosen in a realistic ratio to the bit length $b$. Otherwise, one might be able to increase the total SNR value just by increasing $T_{50}$, which in general is obviously not true. Therefore, we again scale the reader resolution $R$ (which is proportional to $T_{50}$) with the bit length $b$ according to Eq. \eqref{eq::scaling} to keep the factor $T_{50} / b$ in Eq. \eqref{eq::prop} constant. Finally, we note that the reader width $W$ is also kept constant. Under these assumptions, we can write the SNR value as a function of the remaining variables and a proportionality constant $f$ in the form

\begin{align}\label{eq::fitfunction}
\begin{split}
\text{SNR}(f,\sigma_d, D, b) =\quad\quad\quad\quad\quad& \\ f \cdot  \left(\sigma_d^2 + \frac{(D+B)^2}{12} \right) ^{-1} & \cdot (D+B)^{-1}\cdot b^{2}.
\end{split}
\end{align}

\begin{figure}[htbp]
\centerline{\includegraphics[width=0.52\textwidth]{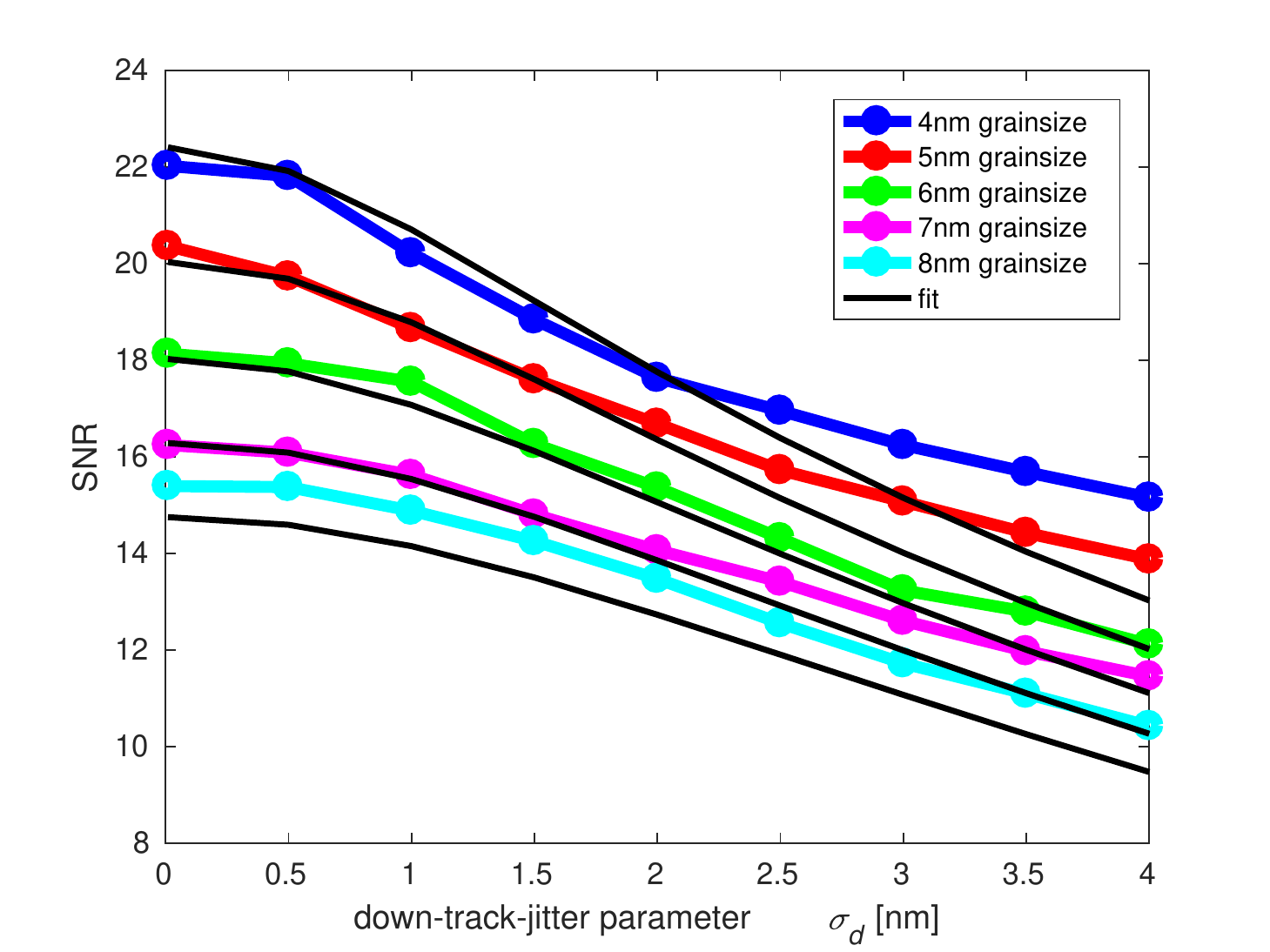}}
\caption{Fitting curves of the SNR calculation for various down-track-jitter parameters and constant bit length $b=10$ nm.}
\label{fig::fit}
\end{figure}

\begin{figure}[htbp]
\centerline{\includegraphics[width=0.52\textwidth]{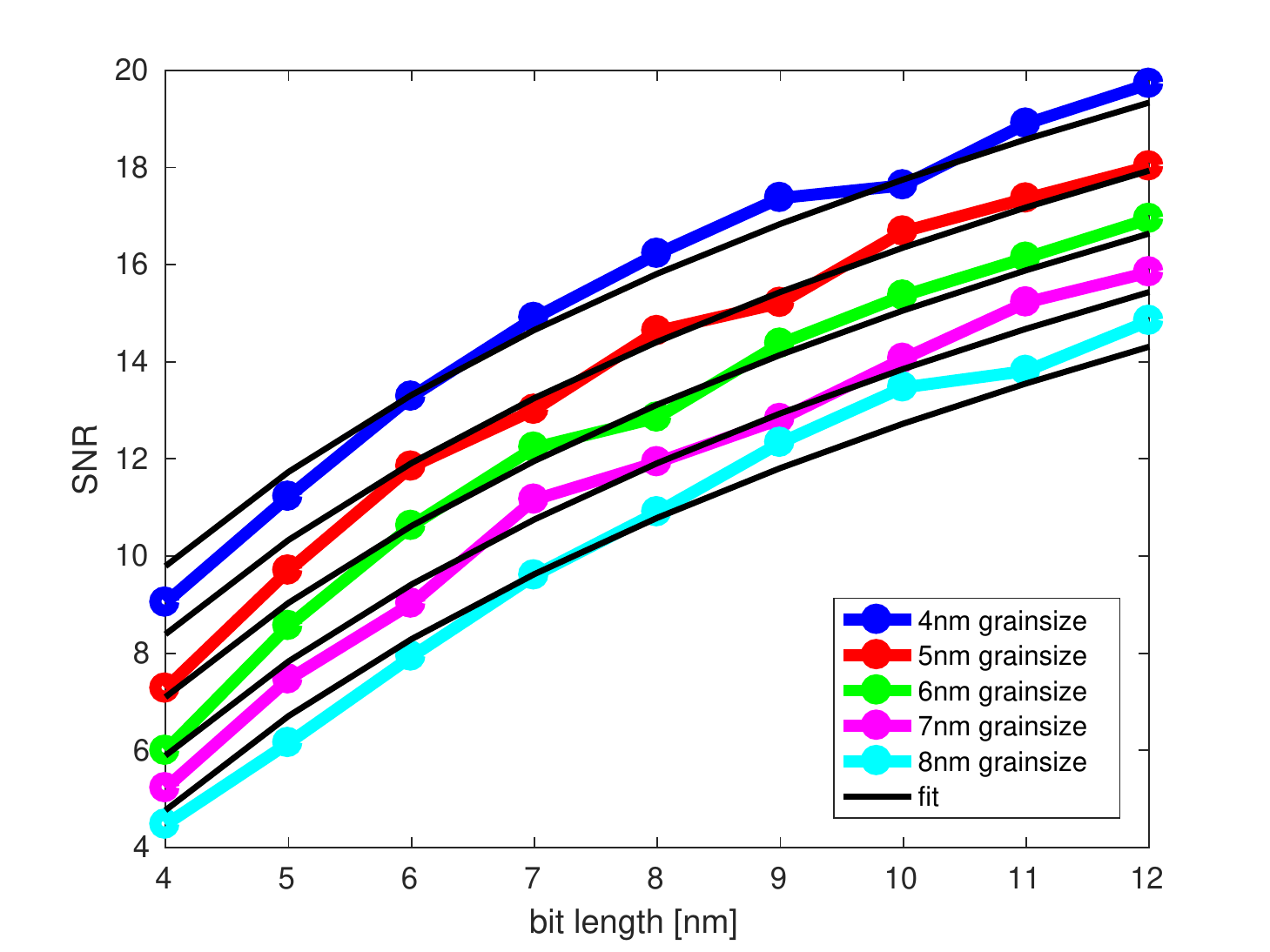}}
\caption{Fitting curves of the SNR calculation for various bit lengths and constant down-track-jitter parameter $\sigma_d=2$ nm.}
\label{fig::fit2}
\end{figure}

For the given values for $\sigma_d\in [0.01,4.00]$~nm, $b\in [4.0,12.0]$~nm, $d\in [4,8]$~nm and the corresponding calculated SNR values, this function can be fitted with the fitting parameter $f$. Note that in contrast to the plots in Fig. \ref{fig::results}, we now choose a uniform basic parameter set for all grain diameters, namely the values for $5$~nm grain diameter in Tab. \ref{tab::fit}. Otherwise, we would additionally have to consider the grain size dependent variation of parameters, which are not included in the model function in Eq. \eqref{eq::fitfunction}. We obtain the value of the fitting parameter $f=18.1246$~nm and the results are shown exemplarily in Fig. \ref{fig::fit} and \ref{fig::fit2}.

The plots show that the trend of the fit agrees quite well with the actual SNR curves, however for $\sigma_d \longrightarrow 4$~nm and $D\longrightarrow 8$~nm we obtain deviations. Taking into account the approximations in both the writing process (neglection of grain size distribution) and the derivation of the formula in Eq. \eqref{eq::fitfunction} (simple model of square grains for Eq. \eqref{eq::model}, no dependency on $P_{\text{max}}$ etc.), those seem acceptable.

\section{Conclusion and Remarks}
We developed an analytical model for the switching probability phase diagram of a magnetic grain in a recording medium during heat-assisted magnetic recording (HAMR). Such a phase diagram describes the switching probability of the grain depending on its down-track and off-track position and is thus very valuable to evaluate the performance of a given recording setup. The proposed model has eight input parameters, determining the bit's position, its dimensions, its jitter in down-track and off-track direction, its curvature and the maximum switching probability in the center of the bit. By mapping the switching probabilities onto a recording medium and calculating the read-back signal with a given sensitivity function it is possible to model the whole HAMR write and read cycle with little computational effort. Additionally, due to the possibility of independent parameter variation in the model, we could investigate the influence of each parameter on the resulting signal-to-noise ratio (SNR) separately. Our results showed the impact of the bit length, curvature, maximum switching probability and down-track-jitter on the final SNR of written bit patterns. Whereas the variation of the bit length, maximum switching probability and down-track-jitter led to differences of $10, 9$ respectively $5$~dB, we could only show a gain of about $1$~dB for the reduction of bit curvature. Furthermore, the comparison with theoretical equations led to good agreement. 

Since there are considerable efforts to optimize the bit quality in HAMR containing material design and writing techniques, our approach of using an analytical model for the whole recording cycle could provide a qualitative a priori indication about the cost-benefit ratio of a recording setup in terms of the SNR with low computational effort. In addition, with the proposed model the effects of individual changes, such as the down-track jitter or the transition curvature, on the SNR can be studied separately. This is not possible with direct simulations of the write process, where only material parameters of the recording grains or write parameters of the recording head can be changed, which then has an impact on many aspects of the resulting footprint.

\section*{Acknowledgment}
The authors would like to thank the Austrian Science Fund (FWF) under grant No. I2214-N20 and the ASRC/IDEMA for financial support. The computational results presented have been achieved using the Vienna Scientific Cluster (VSC). Further thanks to Stephanie Hern\'{a}ndez for providing the SNR calculator from SEAGATE, which we gratefully used for the SNR calculation of the read-back signal and Prof. Randall Victora for providing the sensitivity function of the reader.

\end{document}